\documentclass[a4paper,11pt]{article}
\pdfoutput=1

\usepackage{jheppub}

\usepackage{rotating}
\usepackage{graphicx}
\usepackage{booktabs}
\usepackage{amsmath}
\usepackage{arydshln}
\usepackage{latexsym}
\usepackage{mathrsfs}
\usepackage{amssymb}
\usepackage{multirow}                   
\usepackage{slashed}
\usepackage{bm}
\usepackage{hyperref}
\usepackage{pdflscape}

\usepackage[T1]{fontenc}
\usepackage[dvipsnames,table]{xcolor}   
\usepackage{comment}
\usepackage[normalem]{ulem}
\usepackage{longtable}

\newcommand{\idlr}[1]{i\!\overleftrightarrow{D}_{\!\!#1}}

\usepackage{bbold} 

\title{
Squeezing Proton Decay and Neutrino Masses:\\ Upper Bounds on Standard Model Extensions
}

\author[a,b]{Arnau Bas i Beneito,}
\author[a,b,d]{John Gargalionis,}
\author[a,b]{Juan Herrero-Garc\'{i}a,}
\author[c]{Michael A. Schmidt}

\affiliation[a]{Departament de Física Teòrica, Universitat de València, 46100 Burjassot, Spain}

\affiliation[b]{Instituto de Física Corpuscular (CSIC-Universitat de València),
Parc Científic UV, C/Catedrático José Beltrán, 2, E-46980 Paterna, Spain}

\preprint{IFIC/23-52, CPPC-2023-12}

\affiliation[c]{Sydney Consortium for Particle Physics and Cosmology,
  School of Physics, University of New South Wales,
  Sydney, New South Wales 2052, Australia}

\affiliation[d]{
  University of Adelaide, ARC Centre of Excellence for Dark Matter Particle Physics \& CSSM, Department of Physics, Adelaide SA 5000 Australia
}

\emailAdd{arnau.bas@ific.uv.es}
\emailAdd{john.gargalionis@adelaide.edu.au}
\emailAdd{juan.herrero@ific.uv.es}
\emailAdd{m.schmidt@unsw.edu.au}

\date{\today}
\preprint{CPPC-2025-03}
\abstract{
Baryon and lepton number are excellent low-energy symmetries of the Standard Model (SM) that tightly constrain the form of its extensions. In this paper we investigate the possibility that these accidental symmetries are violated in the deep UV, in such a way that one multiplet necessary for their violation lives at an intermediate energy scale $M$
above the electroweak scale. We write down the simplest effective operators containing each multiplet that may couple linearly to the SM at the renormalisable level and estimate the dominant contribution of the underlying UV model to the pertinent operators in the SMEFT: the dimension-5 Weinberg operator and the baryon-number-violating operators up to dimension 7. Our results are upper bounds on the scale $M$ for each multiplet--operator pair, derived from neutrino-oscillation data as well as prospective nucleon-decay searches. We also analyse the possibility that both processes are simultaneously explained within a natural UV model. In addition, we advocate that our framework provides a convenient and digestible way of organising the space of UV models that violate these symmetries.}


\makeatletter
\gdef\@fpheader{}
\makeatother 
\begin{document} 
\maketitle

\flushbottom

\section{Introduction}\label{sec:introduction}

The Standard Model (SM) of particle physics is one of the most accurate theories describing the interaction of subatomic particles. However, despite its remarkable accuracy, it falls short of providing an adequate explanation of the phenomena of neutrino oscillations and the excess of matter over anti-matter, among others. 
The measured neutrino oscillations~\cite{Super-Kamiokande:1998uiq,KamLAND:2002uet,SNO:2002tuh} point towards the requirement of massive neutrinos, and for the matter-antimatter asymmetry to be dynamically generated in the Universe the Sakharov conditions need to be fulfilled~\cite{Sakharov:1967dj}. Assuming that neutrinos are Majorana particles,\footnote{Whether neutrinos are Dirac or Majorana remains still unknown. Future experiments such as neutrinoless double beta decay ($0\nu\beta\beta$) may shed light on this question~\cite{PhysRev.56.1184} (see Ref.~\cite{DellOro:2016tmg} for a review). From an Effective Field Theory (EFT) point of view, we believe Majorana neutrinos are better motivated and this will be our hypothesis.} the explanation of these phenomena necessarily implies the violation of two accidental symmetries of the SM: Baryon Number ($B$) and Lepton Number ($L$). More precisely, Baryon Number Violation (BNV) and Lepton Number Violation (LNV) must be incorporated into any theory meant to account for these shortcomings of the SM. 
Lepton number is violated in well-motivated neutrino mass models like the so-called seesaw models~\cite{Minkowski:1977sc,Yanagida:1979as,Gell-Mann:1979vob,Yanagida:1979gs,Glashow:1979nm,
Mohapatra:1979ia,Magg:1980ut,Lazarides:1980nt,
Foot:1988aq} (see also Refs.~\cite{Barbieri:1979ag,Wilczek:1979hh,Witten:1979nr,Weinberg:1979sa,Schechter:1980gr,Yanagida:1980xy}), while both baryon and lepton numbers are violated in Grand Unified Theories (GUTs)~\cite{Georgi:1974sy,Fritzsch:1974nn}, whose main prediction at low energies is BNV nucleon decay. 

A suitable way to study whatever new physics (NP) might mediate violations of these symmetries is through the use of Effective Field Theory (EFT). This approach has become popular in recent years, as evidenced by the rise of the so-called SM Effective Field Theory (SMEFT)~\cite{Brivio:2017vri,Grzadkowski:2010es,Buchmuller:1985jz,Henning:2014wua} as a paradigm for interpreting deviations from the SM and for bottom-up model building. In the SMEFT, non-zero neutrino masses are predicted as a leading sign of physics beyond the SM, with $L$ violated in two units at dimension 5 by the Weinberg operator~\cite{Weinberg:1979sa}. Similarly, $B$ is first violated in one unit at dimension 6~\cite{Weinberg:1979sa,Weinberg:1980bf,Wilczek:1979hc,Abbott:1980zj}, and the implied BNV nucleon decays are both the most striking prediction of the dimension-6 Lagrangian as well as their most sensitive probe.

In this context, the smallness of the neutrino masses can be explained by imagining that the dimension-5 operator is suppressed by a very large scale. The simplest, concrete models that realise this possibility are the seesaw models, which generate the Weinberg operator at tree level. Despite their economy, such models push the scale of LNV beyond any currently available experimental probe. Another more testable possibility to explain neutrino masses is to imagine generating the Weinberg operator through loops. Such radiative models have a long history~\cite{Zee:1980ai,Babu:1988ki,Cai:2017jrq}, and the theory-space has been systematically mapped through complementary computational approaches. The first is based on loop-level completions of operators giving rise to neutrino masses at tree level~\cite{Farzan:2012ev,Bonnet:2012kz,AristizabalSierra:2014wal,Cepedello:2018rfh,Cepedello:2017eqf}, which we refer to as the loop-level-matching paradigm; and the second is based on tree-level completions of high-dimensional LNV operators that lead to neutrino masses at loop level~\cite{Babu:2001ex,deGouvea:2007qla,Angel:2012ug,Cai:2014kra,Gargalionis:2020xvt}, an approach we dub the tree-level matching paradigm. The same methods have quite recently also been applied to the study of nucleon decays~\cite{Helo:2019yqp, Gargalionis:2024nij}, with the aim of exploring low-scale models of BNV, and some efforts have been made to explore possible connections between both phenomena within a single UV model~\cite{Guth:1981uk,Chang:1980ey,deGouvea:2014lva,Boucenna:2014dia,DiIura:2016wbx,Hagedorn:2016dze,Nomura:2024zca,Kang:2024oyf}.

A powerful way to parametrise patterns of tree-level deviation from the SM is to study the exotic multiplets that couple linearly to the SM at the renormalisable level. These Linear SM Extensions (LSMEs) were introduced systematically in Ref.~\cite{deBlas:2017xtg} as the collection of exotics fields that couple linearly to the SM via renormalisable operators. They have many practical uses; for example, in making complex fits of SMEFT coefficients to experimental data more tractable~\cite{Ellis:2020unq,Gargalionis:2024jaw,terHoeve:2025gey}, and in classifying the space of neutrino-mass models in the tradition of the tree-level matching paradigm~\cite{Herrero-Garcia:2019czj}. 

In this paper we aim to constrain the LSMEs by analysing their contributions to LNV and BNV phenomena using EFT. The analysis applies to the simplest and most minimal UV models in which the LSME appears. To achieve this, we write down effective operators that include such exotic multiplets. We identify the pertinent $\Delta L = 2$ and $\Delta B = 1$ operators in each case and estimate the dominant contribution to both neutrino masses and BNV nucleon decay in the underlying UV models implied by these operators. Upper bounds on the mass of each LSME can also be set by demanding that the atmospheric bound on the mass of the heaviest neutrino derived from the atmospheric mass squared difference $\Delta m_{\rm atm}^2$ is reproduced, assuming that the combination of couplings we identify for each LSME is the dominant source of LNV. Similarly, we can set an upper bound by imagining that a positive signal is seen at the next generation of nucleon-decay experiments, mediated by the LSME being studied and assuming that it couples in the way we identify (see also Ref.~\cite{Bermejo2022}).\footnote{Similarly, lower bounds can be easily obtained, for a given set of couplings, by rescaling the upper limits provided with current null-results for nucleon decay searches and with the upper limit on the sum of neutrino masses from cosmology and tritium $\beta-$decay.} This analysis is motivated by the increased sensitivity expected at the next generation of neutrino-oscillation and nucleon-decay experiments, such as Hyper-Kamiokande (Hyper-K)~\cite{Hyper-Kamiokande:2022smq}, Deep Underground Neutrino Experiment (DUNE)~\cite{DUNE:2020ypp}, Jiangmen Underground Neutrino Observatory (JUNO)~\cite{JUNO:2015zny}, and THEIA~\cite{Theia:2019non}.

As a by-product of our analysis, we advocate that our framework can also be seen as a classification system for the UV models derived in the aforementioned tree-level-matching paradigm. Since each tree-level completion of a SMEFT operator must contain at least one LSME, the EFTs we describe can stand in for families of UV models in which the lightest LSME is partially resolved. Viewed in this way, it is a useful half-way point between the full UV models, whose number in the case of neutrino masses is potentially in the tens of thousands~\cite{Gargalionis:2020xvt}, and the SMEFT, where no model-specific information is present.

The rest of this paper is organised as follows. In Sec.~\ref{sec:methods} we introduce the list of 48 LSMEs following the convention of Ref.~\cite{deBlas:2017xtg}\footnote{A different convention was used in Ref.~\cite{Herrero-Garcia:2019czj}. For instance, we identify $\Delta \sim \Xi_1$, $h \sim \mathcal{S}_1$, $L_2 \sim \Delta_3$, etc.} and the procedure we follow to find bounds on the masses of the UV particles coming from neutrino masses and nucleon decay. In Sec.~\ref{sec:results} we show the operator that gives the dominant contribution to such processes for each field as well as the derived bounds on the masses of the multiplets. In Sec.~\ref{sec:conclusions} we summarise the main conclusions of our work and comment on possible follow-up ideas. The appendices contain additional technical details.

\section{Methodology}\label{sec:methods}

Below we outline the methods we use to estimate the dominant contributions to the neutrino masses and BNV nucleon decays. This includes a description of the EFT framework we work in, our approach to the setting of upper bounds on the masses of each exotic particle, and the key assumptions that underlie these limits.

\subsection{Theoretical framework}\label{sec:framework}

Our goal is to explore how each of the LSMEs might be responsible for the (potentially simultaneous) generation of neutrino masses and nucleon decays. We do this by working in an EFT framework in which one of these multiplets extends the SMEFT. There are 48 multiplets that couple linearly to the SM at the renormalisable level: 19 scalars, 13 fermions and 16 vectors. These multiplets generate dimension-5 and dimension-6 operators in the SMEFT at tree level, and for this reason the most adopted convention for their nomenclature comes from the dimension-6 tree-level UV/IR dictionary, the so-called Granada dictionary~\cite{deBlas:2017xtg}, whose notation we also adopt here. We work with the Lorentz vectors as Proca fields and do not specify further how they obtain their mass. Scalars associated with the mass mechanism are chosen to be heavier than the corresponding Lorentz vector. In addition, we work with spinors as vector-like Dirac fermions or Majorana fermions.

\begin{figure*}[ht]
    \centering
    \includegraphics[width=0.35\textwidth]{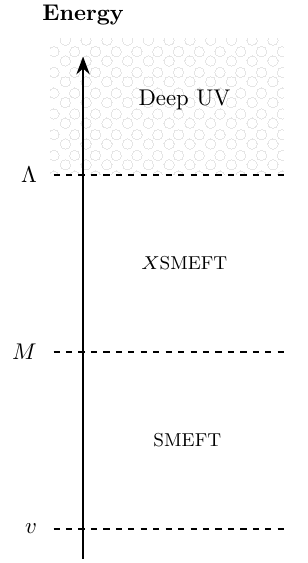}
    \includegraphics[width=0.63\textwidth]{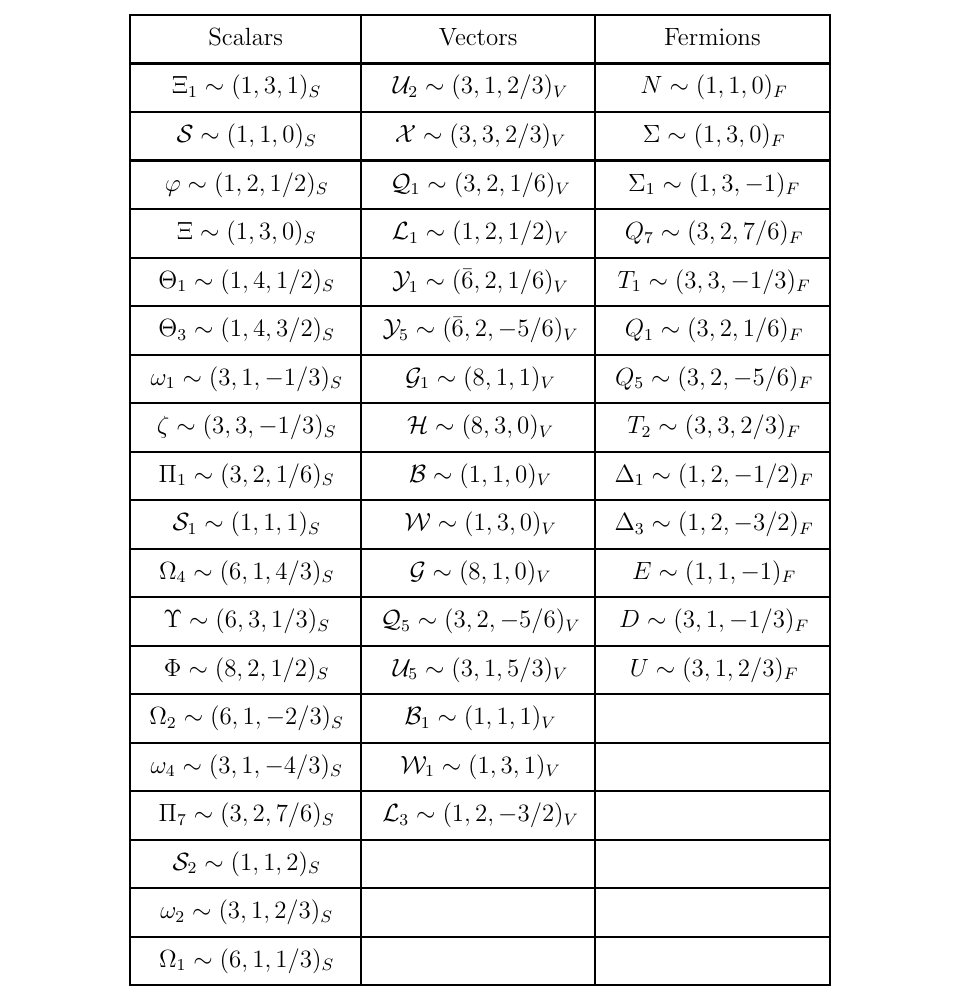}
    \caption{(Left) An illustration of the tower of EFTs we use in our analysis. The full theory describing the violation of baryon and lepton number lives above the scale $\Lambda$, while the intermediate scale $M$ characterises one of the LSMEs participating in the neutrino-mass or nucleon-decay mechanism. (Right) Quantum numbers of the 48 LSMEs analysed in this work under the Lorentz and gauge group $G_{\rm SM} =\left(\rm{SU}(3)_C, \, \rm{SU}(2)_L, U(1)_Y \right)$. Each particle $X$ is assumed to exist within the $X$SMEFT regime, while the unknown UV theory lives above $\Lambda$.}\label{fig:energyscalesandparticles}
\end{figure*}

We imagine that a UV-complete Majorana neutrino-mass model or nucleon-decay mechanism involving at least one LSME operates above the scale $\Lambda$, a regime we call the \textit{deep UV}. Further, we assume that one of the LSMEs participating in this mechanism carries a mass $M$, potentially much smaller than $\Lambda$ but larger than the electroweak scale.
We extend SMEFT with this exotic multiplet and call each such EFT the $X$SMEFT,\footnote{Such EFTs have been called BSMEFTs in Refs.~\cite{Adhikari:2020vqo,Banerjee:2020jun}. Our use of $X$SMEFT here differs from the general concept of BSMEFT since we specifically focus on single-multiplet additions to the SM field content by one of the LSMEs. They are often denoted by the additional LSME prefixing the `SMEFT' label, as in $\nu$SMEFT~\cite{delAguila:2008ir,Liao:2016qyd}, for instance.} where $X$ generically represents one of the 48 multiplets constituting the linear extensions of the SM, \textit{i.e.}\ $N$SMEFT, $\omega_1$SMEFT, \textit{etc.} When matching the underlying UV model to the $X$SMEFT at the scale $\Lambda$, BNV and LNV operators without the field $X$ are also generated. Their Wilson coefficients can be estimated from the $X$SMEFT operator through naive dimensional analysis following arguments similar to those of Refs.~\cite{deGouvea:2007qla,Gargalionis:2020xvt,Gargalionis:2024nij}. The SMEFT coefficients are then used to estimate the contribution to neutrino masses and the rates of BNV nucleon decays, from which we can derive approximate bounds on combinations of $X$SMEFT parameters, including the scale $M$. These energy regimes and the two different EFTs we consider are depicted pictorially in Fig.~\ref{fig:energyscalesandparticles}, along with the definition of the 48 LSMEs analysed in this work.

Our aim is to estimate the dominant contributions from the unknown UV mechanism living above $\Lambda$ to the appropriate SMEFT operators. In the case of neutrino masses this is the dimension-5 Weinberg operator~\cite{Weinberg:1979sa}
\begin{equation} \label{eq:weinbergop}
    \left[\mathcal{O}_{5}\right]_{pq} =(L_p^iL_q^j)H^kH^l\epsilon_{ik}\epsilon_{jl} \;,
\end{equation}
while the largest contributions to two-body nucleon decays come from $\Delta B = 1$ operators entering at either dimension 6 or dimension 7, based on symmetry grounds: dimension 6 for those that conserve $B-L$, and dimension 7 for those that violate $B-L$ in two units.

The $\Delta B = \Delta L = 1$ SMEFT operators of dimension 6 are defined as~\cite{Weinberg:1979sa,Grzadkowski:2010es}\footnote{Here and throughout this work we employ two-component Weyl spinor notation. We follow the conventions of Ref.~\cite{Dreiner:2008tw}, including those for the SM fermions defined in App.~J in Ref.~\cite{Dreiner:2008tw}.}
\begin{equation}
  \label{eq:smeft-ops-1}
  \begin{aligned}
    \left[\mathcal{O}_{qqql}\right]_{pqrs} & = (Q_p^{i} Q_q^{j}) (Q_r^{l}L_s^k )\epsilon_{ik} \epsilon_{jl}\,,
    &
    \left[\mathcal{O}_{qque}\right]_{pqrs} & = (Q^{i}_{p} Q^{j}_{q}) (\bar u_r^{\dagger}\bar e_s^\dagger) \epsilon_{ij}\,,
    \\
    \left[\mathcal{O}_{duue}\right]_{pqrs} & =  ( \bar d_p^{\dagger} \bar u_q^{\dagger}  )(\bar u_r^{\dagger}\bar e_s^\dagger )\,,
    &
    \left[\mathcal{O}_{duql}\right]_{pqrs} & = (\bar d_p^{\dagger}\bar u_q^{\dagger} ) ( Q_r^{i} L_s^j) \epsilon_{ij}\,.
  \end{aligned}
\end{equation}
where $i, j,k,l$ are SU(2)$_{\rm L}$ indices. Colour indices are suppressed here and in the following and it is understood that the colour indices of the three quarks in each bilinear are contracted using a Levi-Civita tensor with indices in the order of the quarks in the operators. The $\Delta B = -\Delta L = 1$ SMEFT operators of dimension 7 are defined as~\cite{Liao:2020zyx}
\begin{equation}
  \label{eq:smeft-ops-2}
  \begin{aligned}
    \left[\mathcal{O}_{\bar l dddH}\right]_{pqrs} & = (L_p^{\dagger} \bar d_q^{\dagger}) (\bar d^{\dagger}_{r} \bar d^{\dagger}_{s}) H\,,
    &
    \left[\mathcal{O}_{\bar ldqq \tilde H}\right]_{pqrs} & = (L_p^{\dagger} \bar d_q^{\dagger}) (Q_r Q_s^{i} ) \tilde H^j  \epsilon_{ij}\,,
    \\
    \left[\mathcal{O}_{\bar e qdd\tilde H}\right]_{pqrs} & = (\bar e_p Q_q^{i})(\bar d_r^{\dagger} \bar d_s^{\dagger}) \tilde H^j \epsilon_{ij}\,,
    &
    \left[\mathcal{O}_{\bar ldud\tilde H}\right]_{pqrs} & = (L_p^{\dagger} \bar d_q^{\dagger})(\bar u^{\dagger}_r \bar d^{\dagger}_s) \tilde H\,,
    \\
    \left[\mathcal{O}_{\bar lqdDd}\right]_{pqrs} & = (L^{\dagger}_p \bar \sigma^\mu Q_q) ( \bar d_r^{\dagger} \idlr{\mu} \bar d_s^{\dagger})\,,
    &
    \left[\mathcal{O}_{\bar edddD}\right]_{pqrs} & = (\bar e_p \sigma^\mu \bar d_q^{\dagger}) ( \bar d_r^{\dagger} \idlr{\mu} \bar d_s^{\dagger})
    \;.
  \end{aligned}
\end{equation}

Importantly, our estimates assume that the essential elements of the full-theory description of nucleon decays or neutrino masses are captured by the $X$SMEFT operator chosen and the renormalisable couplings of $X$ to the SM. We use $\mathcal{O}_{X}$ for the $X$SMEFT operator and $c$ for its dimensionless Wilson coefficient. Additionally, we label $y$ the coupling constant of the renormalisable interaction of $X$ to the SM appearing in the dominant contribution to nucleon decays or neutrino masses in the presence of $\mathcal{O}_{X}$. An equivalent phrasing of the aforementioned condition is then that the largest contribution to nucleon decays or neutrino masses is directly proportional to the product $c\,y$. The choice of the renormalisable operator is generally fixed by the quantum numbers of the LSME,\footnote{In some cases, more than one renormalisable operator is allowed, such as for $\mathcal{B}$, $\mathcal{W}$, $\mathcal{G}$, $\varphi$, $\omega_1$, $\omega_4$, $\zeta$, $\mathcal{Q}_1$, $\mathcal{Q}_5$, $N$, $\Sigma$, and $\Xi_1$. For such cases, we choose the one that leads to the largest contribution to either the Weinberg operator or the $\Delta B = 1$ operators at $d \leq 7$.} and we choose to study $X$SMEFT operators of the lowest mass dimension whose combined action violates the appropriate symmetries, \textit{i.e.}\ $\Delta L=2$ or $\Delta B=1$. In case there are multiple effective operators entering at the same mass dimension, we choose the one that provides the largest contribution to either the Weinberg operator or the $\Delta B = 1$ operators at $d \leq 7$. For more details on this choice see Sec.~\ref{sec:genuineness}.

We distinguish two ways in which the Weinberg operator or the $d \leq 7$ baryon-number-violating operators may arise at the low scale:
\begin{enumerate}
\item Renormalisation group mixing of low-dimensional lepton- and baryon-number-violating operators in the $X$SMEFT featuring the exotic field $X$ into the appropriate SMEFT operators between the scales $\Lambda$ and $M$;
\item Loop-level matching at the scale $\Lambda$ onto the relevant SMEFT operators.
\end{enumerate}
The dominant mechanism depends on the new field $X$ and the dimension of the new lepton- and baryon-number-violating operators. For this reason it is useful to define $\Delta d \equiv d_{X} - d$, where $d$ stands for the mass dimension of the SMEFT operator, and $d_X$ that of the $X$SMEFT operator. In the following, neutrino masses (BNV nucleon decays) are characterised by $d=5$ ($d=6,7$) contributions.

For $X$SMEFT operators of the same dimension as the Weinberg or baryon-number-violating operators driving the nucleon decay, \textit{i.e.}\ in cases where $\Delta d = 0$, renormalisation group mixing generally dominates, since it is enhanced by factors of logarithms of $\Lambda/M$. These contributions can be calculated directly in the $X$SMEFT:
\begin{equation}
  \label{eq:deltad-0}
  C_{\text{SMEFT}} \sim  
  \left(\frac{ \log \left(\frac{\Lambda}{M}\right) }{16\pi^2} \right)^\ell 
  \frac{c\, y}{\Lambda^{d-4}} 
  \prod_i g_i
  \; ,
\end{equation}
where $g_i$ are the SM gauge and Yukawa couplings, the index $i$ depends on the multiplet-pair operator considered, and $\ell$ is the number of loops.

The $\Delta d = 1$ operators in the $X$SMEFT may still generate the dominant contribution to the SMEFT operator for fermions $X$ if the contribution is proportional to the fermion mass $M$, 
\begin{equation}
  \label{eq:deltad-1a}
C_{\text{SMEFT}} \sim 
  \left(\frac{ \log \left(\frac{\Lambda}{M}\right) }{16\pi^2} \right)^\ell 
\frac{M}{\Lambda} \frac{c \,y}{\Lambda^{d-5}} 
\prod_i g_i \; ,
\end{equation}
or for scalars $X$ with a trilinear interaction $\mu$
\begin{equation}
  \label{eq:deltad-1}
C_{\text{SMEFT}} \sim 
  \left(\frac{ \log \left(\frac{\Lambda}{M}\right) }{16\pi^2} \right)^\ell 
\frac{\mu}{\Lambda} \frac{c}{\Lambda^{d-5}} 
\prod_i g_i \; ,
\end{equation}
where we have explicitly included the case of $\ell$-loop mixing into the pertinent SMEFT operator. Here, we highlight the suppression coming from the explicit factor of $M/\Lambda$ or $\mu/\Lambda$.

For $\Delta d > 1$ $X$SMEFT operators, we expect that loop-level matching at the scale $\Lambda$ dominates. The power counting implied by the mass-dimension of the $X$SMEFT operator is illusory in this case, because the calculation is carried out in the full theory and not in $X$SMEFT. Taking for concreteness the example of neutrino masses, all of the implied neutrino masses scale as $v^2/\Lambda$, since they generate the dimension-5 Weinberg operator at the scale $\Lambda$. There is, however, a general tendency for models generating operators of large mass dimension at tree level to generate smaller matching estimates~\cite{Gargalionis:2020xvt}. Additionally, in cases where models generating high-dimensional operators at tree level dominate the neutrino masses, this dominance is abruptly undermined by even a small departure of the UV couplings below unity. Further, the models that generate such operators become increasingly baroque as the mass dimension of the $X$SMEFT operator increases. In these cases we proceed by using the known results of Refs.~\cite{deGouvea:2007qla,Gargalionis:2020xvt,Gargalionis:2024nij} to provide a rough estimate of the matching contribution from the full theory to set conservative bounds on $M$ and $\Lambda$. Loop-level matching at the scale $M$ is subdominant, because it is suppressed by $(M/\Lambda)^{\Delta d}$ and thus is neglected in our study.

We point the reader to App.~\ref{sec:matching-and-running-estimates} for more details related to our estimates of these matching and running contributions, including symmetry-based arguments for the scalings presented in Eqs.~\eqref{eq:deltad-0} --~\eqref{eq:deltad-1}. Additionally,  in App.~\ref{sec:appendixModel} we provide examples in the context of full UV models for each case. Below we present our procedure for the setting of bounds in this framework.

\subsubsection*{Derivation of the limits}

In order to place an upper bound on the scale characterising any new physics, it is necessary to have a positive experimental signature for departure from the SM. In the case of neutrino masses, we already have a plethora of such measurements. In the case of BNV, we \textit{assume} that a positive signal is observed at Hyper-K, and derive limits based on this assumption. The procedure for deriving the upper bounds presented in this work is described in detail below. We begin with general comments, and then discuss the individual cases of neutrino masses and BNV nucleon decays.

All of the upper bounds we present are estimated on the basis of the central assumption we introduced above: that the dominant contribution to either neutrino masses or nucleon decay involves the operator $\mathcal{O}_X$ along with one of the renormalisable interactions of $X$ to the SM. Since we are setting an upper limit, for simplicity we choose to saturate the inequality $M < \Lambda$; that is, we take $M \sim \Lambda$ and set a conservative bound on $\Lambda$ from the matching estimate derived assuming the limiting case of one UV scale. This procedure is described in more detail below. To set a numerical upper limit, we saturate the perturbativity bounds at unity for the dimensionless couplings ($c$ and $y$ in our notation) and the dimensionless combination $\mu/M$ (the ratio of a trilinear coupling to the intermediate mass scale).\footnote{The choice to take $\mu/M \lesssim 1$ can be motivated in number of ways. See Ref.~\cite{Herrero-Garcia:2014hfa} for a discussion.} We highlight that our choices for the free, exotic parameters are flavour blind. Such a democratic assignment in each model implies that expressions for the entries of the neutrino-mass matrix will inherit the dominance of the third-generation SM Yukawa couplings, where such Yukawa matrices appear in the expressions.

The matching and running estimates are computed by naive dimensional analysis; for each of the combinations of operators that violate $L$ or $B$, we close off the $X$SMEFT operator $\mathcal{O}_X$ with an insertion of the renormalisable interaction identified to provide the dominant mechanism for neutrino masses or nucleon decay in order to form either the Weinberg operator or one of the dimension-6/7 $\Delta B = \pm \Delta L = 1$ nucleon-decay-inducing operators, respectively. To emphasise that these are estimates, we round the numerical values appearing in the limits to either $1\cdot10^n$ or $5\cdot10^n$, where $n$ is an integer representing the order of magnitude.

For the upper bounds derived from neutrino masses we follow the methods of Refs.~\cite{deGouvea:2007qla,Gargalionis:2020xvt,Herrero-Garcia:2019czj} closely. Our central assumption enforces that the atmospheric bound on the mass of the heaviest neutrino $m_\nu > \sqrt{\Delta m^2_{\rm{atm}}} \simeq 0.05$~eV is explained by the interactions chosen for each multiplet $X$. We apply this bound directly on each expression for $m_\nu$ we estimate in order to derive an approximate upper bound on $M$.

As discussed briefly above, for the case of nucleon decay we \textit{assume} that a positive BNV nucleon decay signal is observed at Hyper-K. To derive a bound from two-body nucleon decays we follow Ref.~\cite{Gargalionis:2024nij} and work from the results presented there. In particular, one can write the nucleon decay rate as
\begin{equation}
    \Gamma(N\to M+\ell) = \frac{m_N}{32\pi}\left( 1 - \frac{m_M^2}{m_N^2}\right)^2 \left| \sum_I C_I W_0^I (N \to M) \right|^2 \, ,
\end{equation}
where $m_N$ ($m_M$) are the neutron (meson) mass, and we have neglected the lepton mass in the final state. Here $W_0^I$ is the nuclear matrix elements, which we take from Ref.~\cite{Yoo:2021gql}, and $C_I$ denote the BNV LEFT WCs, which in our setup are generated at tree-level or at loop-level from the dimension-6 or dimension-7 BNV operators listed in Eqs.~\eqref{eq:smeft-ops-1} and~\eqref{eq:smeft-ops-2}. When nuclear matrix elements are unavailable, we use naive dimensional analysis as described in Ref.~\cite{Gargalionis:2024nij}.
We assume that the dominant signature at Hyper-K identified by Ref.~\cite{Gargalionis:2024nij} is observed for each combination of renormalisable and non-renormalisable operator involving $X$, such that a BNV nucleon decay rate is measured within some experimental error, which we ignore. In this case, an upper bound on $M$ can be placed by imposing the aforementioned perturbativity bound on the unknown exotic couplings in the decay rate.

\subsection{Genuineness procedure} \label{sec:genuineness}

In this work, we quote upper bounds from estimating the tree- or loop-matching of the $X$SMEFT operators of the tables of App.~\ref{sec:appendixtables} onto SMEFT operators. It is essential to verify that these operators consistently provide the dominant contribution to Majorana neutrino masses or nucleon decay. This is a non-trivial consistency check, as the models that UV complete the $X$SMEFT operators may include a subset of particles that gives rise to the same phenomenon more dominantly. In such cases, the upper bounds we quote would no longer be meaningful, as they rely on the assumption of being the dominant contribution to the phenomena under study, and thus we must discard the corresponding operator in favour of a different one.

This occurs frequently in our analysis, and if not properly accounted for, one may mistakenly identify the lowest-dimensional operator, typically expected to be the dominant contribution to the phenomena of interest, as the leading effect. Consider, for instance, the active role of $\Theta_1 \sim  (1, 4, 1/2)_S$ in the generation of Majorana neutrino masses radiatively through $\Delta L=2$ operators. A naive approach would suggest that the dominant contribution is given by the dimension-5 $\Theta_1$SMEFT operator $\Theta_{1 \,ijk} L^iL^jH^k$. However, any tree-level UV completion of this operator necessarily involves the type-II or type-III seesaw mediators, $\Xi_1 \sim  (1, 3, 1)_S$ or $\Sigma \sim  (1, 3, 0)_F$, which inherently generate Majorana neutrino masses at tree level, making their contribution more dominant than the loop-induced process.\footnote{We assume the trilinear coupling $\mu \, \Xi_1^\dagger HH$ to be present. While it could be suppressed by a symmetry or taken small as a soft-breaking term, doing so requires additional UV assumptions, which we do not impose.} Therefore, this operator must be disregarded in favour of a higher-dimensional one in the $\Theta_1$SMEFT, specifically at dimension 7, leading to a more suppressed contribution to neutrino mass and thus a more stringent bound on the scale $\Lambda_{\Delta L=2}$.

\begin{figure}[t]
    \centering
    \includegraphics[width=\linewidth]{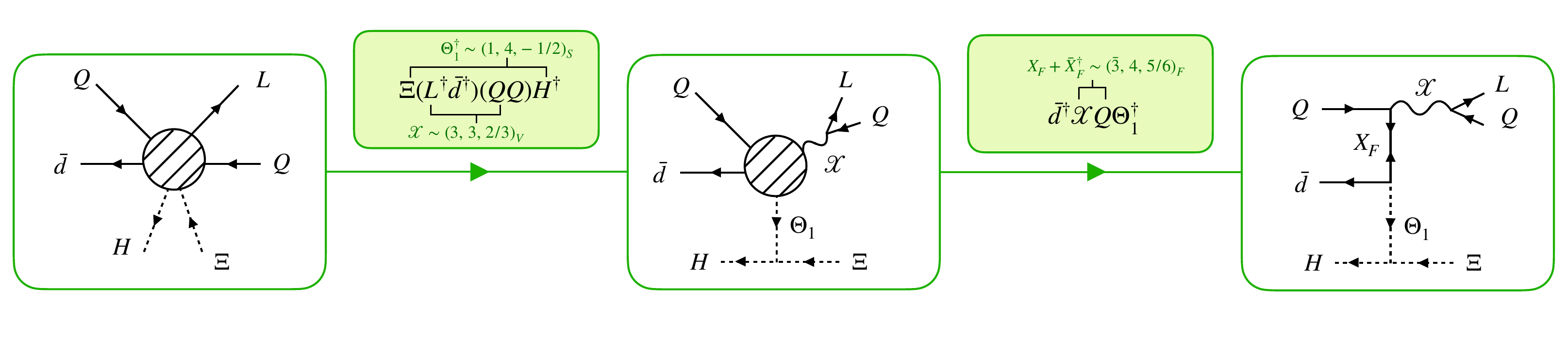}
    \caption{Tree-level completion of operator $\Xi (L^\dagger \bar d^\dagger) (QQ) H^\dagger$. See the main text for details.}
    \label{fig:genuineBNVoperatorExample}
\end{figure}

A similar situation arises for $\mathcal{S} \sim  (\mathbf{1}, \mathbf{1}, 0)_S$, where the lowest-dimensional $\Delta L=2$ and $\Delta B=1$ $\mathcal{S}$SMEFT operators are obtained by appending this singlet scalar to the corresponding lowest-dimensional $\Delta L=2$ and $\Delta B=1$ SMEFT operators. In such cases, it is evident that any UV completion of the $\mathcal{S}$SMEFT operator will involve the same exotic multiplets that generate the SMEFT operator, but with a more dominant contribution. This argument holds for any $\Delta L=2$ and $\Delta B =1$ $\mathcal{S}$SMEFT operator, leading to the conclusion that such a multiplet is effectively excluded from our analysis.\footnote{An alternative approach, not explored in this study, is to assume that the VEV of $\mathcal{S}$ breaks a linear combination of $B$ and $L$. In such a scenario, $\mathcal{S}$SMEFT operators could provide the dominant contribution to these phenomena due to selection rules and symmetry-protected mechanisms~\cite{Chikashige:1980ui,Chikashige:1980qk,Gelmini:1980re,Barbieri:1981yr,Mohapatra:1982xz,Berezhiani:2015afa}. However, this approach relies entirely on assumptions about the UV completion, which lies beyond the scope of this work.} We note that this argument only partially extends to the lowest-dimensional $\Delta B = 1$ $\Xi$SMEFT operators $\Xi (QQ)(QL)$, $\Xi (QQ)(\bar u^\dagger \bar e^\dagger)$, and $\Xi (\bar d^\dagger \bar u^\dagger) (QL)$. In this case, none of these three dimension-7 $\Xi$SMEFT operators provide the dominant contribution to nucleon decay in our setup, as their tree-level completions always include the linear SM extensions that generate dimension-6 $\Delta B = 1$ operators in the SMEFT at tree-level, namely the leptoquarks $\omega_1$, $\omega_4$, $\zeta$, $\mathcal{Q}_1$, and $\mathcal{Q}_5$. However, when proceeding to the next order in mass dimension, genuinely dominant UV mechanisms for nucleon decay can be found in the tree-level completions of dimension-8 $\Xi$SMEFT operators  by selecting specific $\mathrm{SU}(2)_{\rm{L}}$ structures. Consider the operator $\Xi_{ij} (L^\dagger_k \bar d^\dagger) (Q^i Q^j) H^\dagger_l \epsilon^{kl}$, with a definite SU(2)$_{\rm{L}}$ contraction. This operator admits a tree-level UV completion involving the BSM multiplets $\mathcal{X}$, $\Theta_1$, and the non-LSME vector-like fermion $X_F \sim ( \bar{\mathbf{3}}, \mathbf{4}, 5/6)_F$, as illustrated in Fig.~\ref{fig:genuineBNVoperatorExample}, and this model does not necessarily entail another, more dominant contribution to nucleon decay. In fact, this represents the only such UV completion in this case, as it necessarily requires the $\mathrm{SU}(2)_{\rm{L}}$ structure to be fixed.

The technical criterion described above is commonly referred to in the literature as \emph{genuineness}, and the procedure for identifying genuine operators has been dubbed \emph{filtering}~\cite{Bonnet:2012kz,AristizabalSierra:2014wal,Cepedello:2018rfh,Farzan:2012ev}. While this programme has been extensively applied to $\Delta L=2$ SMEFT operators~\cite{Gargalionis:2020xvt}, the analogous one for $\Delta B=1$ SMEFT operators is still lacking. In this work, we do not aim to provide a comprehensive filtering analysis for the BNV SMEFT operators. Instead, we focus on ensuring that the operators that we identify as implying loop-level nucleon-decay mechanisms do not necessarily imply the presence of tree-level proton or neutron decay. This is done by removing the operators whose tree-level completions involve at least one of the five leptoquarks $\omega_1$, $\omega_4$, $\zeta$, $\mathcal{Q}_1$, and $\mathcal{Q}_5$, which generate dimension-6 BNV SMEFT operators at tree level, or the pairs of multiplets identified in recent studies to generate dimension-7 BNV SMEFT operators at tree level~\cite{Li:2022abx,Li:2023cwy}.\footnote{For instance, $\omega_2$ on its own does not induce nucleon decay, but when the same model also includes the vector-like fermion $Q_1$ both dimension-7 operators $\mathcal{O}_{\bar l dud \tilde H}$ and $\mathcal{O}_{\bar l ddd H}$ are generated~\cite{Beneito:2023xbk}.} In this manner, we implement a simplified \emph{genuineness} programme for the $\Delta B=1$ operators relevant to this study.

We are aware of the limitations and caveats of this simplified implementation, and we therefore emphasise a word of caution: in principle, the dimension-6 and -7 SMEFT $\Delta B=1$ operators may be generated by loops involving a subset of the particles that UV complete a given $X$SMEFT operator, potentially leading to a larger contribution to the loop-level matching onto the aforementioned dimension-6 and -7 BNV SMEFT operators. Unlike for the $\Delta L=2$ database~\cite{Gargalionis:2020xvt}, a complete database for $\Delta B=1$ is not yet available in the literature, rendering our analysis insensitive to the possibility described above.\footnote{We refer the interested reader to Ref.~\cite{Gargalionis:2020xvt}, which provides a detailed analysis of \emph{genuineness} for $\Delta L=2$ operators. The main principles outlined there directly apply to the study of $\Delta B = 1$ operators considered in this work. The reference also includes a comprehensive discussion of cases involving multiplets transforming with the same quantum numbers as SM fields, such as $\varphi$, $E$, $U$, $D$, $Q_1$, and $\Delta_1$.} Fortunately, since we focus on \emph{simple} models, specifically those involving a small number of exotic multiplets, corresponding to tree-level completions of low-dimensional $X$SMEFT operators, we note that such cases are rare. These contributions can only arise at higher dimensions in the $X$SMEFT, requiring \emph{at least} two additional BSM multiplets besides the LSME $X$ under study. Such scenarios can only occur for $X$SMEFT operators of dimension $d \geq 7$, so special attention must be paid to these cases.

In addition to extending the filtering procedure described above to the $\Delta B = 1$ operators, we also use an improved version of the original code from Ref.~\cite{Gargalionis:2020xvt} for the $\Delta L=2$ case, as this work also includes vector completions, which were absent in the original study. Finally, it is important to note that the main result of implementing the genuineness programme is the derivation of bounds on the UV scale that are equal or more stringent than those previously established without the genuineness programme in the limiting case $M \to \Lambda$.

\subsection{Antisymmetry in flavour space} \label{sec:flavoursymmetries}

The bounds we supply on BNV-operator coefficients are derived from the database of matching estimates published with Ref.~\cite{Gargalionis:2024nij}. That database presents limits on each operator with as few assumptions as possible made on its flavour, gauge and Lorentz structure, with essentially only the field content fixed.

In the present work, we identify several multiplets for which flavour-index permutation symmetries can be reliably inferred from the non-renormalisable operator, as the renormalisable interaction term imposes additional constraints on its structure. We denote these symmetries using curly (square) brackets to indicate pairs of flavour indices that are (anti)symmetric under permutation. This can impact the strongest constraints on the operator, as the previously most stringent bound may become irrelevant if the corresponding operator vanishes identically. In certain cases involving three identical SM particles or pairs of identical SM particles, the corresponding flavour-index permutation symmetry becomes more intricate, as checked with the \textsc{Mathematica} package \texttt{Sym2Int}~\cite{Fonseca:2017lem,Fonseca:2019yya}. In particular, such couplings appear in the $\Delta L=2$ operators of the LSMEs $\mathcal{Q}_5$, $\mathcal{U}_5$, $\Delta_3$, $\omega_4$, $\Omega_4$, $\Omega_2$, $\Omega_1$, $\mathcal{S}_2$ and $\omega_2$, as well as in the $\Delta B=1$ operators of the LSMEs $\mathcal{S}_2$, $\mathcal{G}_1$ and $\mathcal{B}_1$, so that the reader is aware of these intricacies when analysing their phenomenology.

For example, the multiplet $\Xi_1 \sim  (1, 3, 1)_S$ couples to $H^\dagger H^\dagger$ through a trilinear coupling, and the lowest-dimensional genuine (see Sec.~\ref{sec:genuineness}) $\Delta B=1$ operator constructed from $\Xi_1$ is $\Xi_1 (LQ)( \bar{d}^\dagger \bar{d}^\dagger)$. Both the gauge and Lorentz structure of this operator are uniquely fixed and this results in a specific antisymmetry on the flavour indices of the $( \bar{d}^\dagger \bar{d}^\dagger)$ bilinear. Throughout this work we use the notation of Ref.~\cite{Gargalionis:2024nij} to denote the BNV SMEFT operators. In this notation, the flavour component of $\mathcal{O}_{16} = LQ\bar d^\dagger \bar d^\dagger HH$ that will be most constrained at Hyper-K from the sensitivity information that we have is $1133$~\cite{Gargalionis:2024nij}. This operator vanishes identically in our case, and instead, the best limit is set by the $1132$, for which the resulting limit is further suppressed by a factor of $\sqrt{|V_{ts}^*|}$, where $V$ is the CKM matrix.

A qualitatively different example is provided by $\omega_4$, which at tree level generates the operator $(\bar{u}_{[p} \bar{u}_{q]})(\bar{d}_r \bar{e}_s)$, antisymmetric in the up-type flavour indices. Here, the flavour structure that would realise tree-level nucleon decay vanishes identically, and instead the leading-order contribution comes from the loop-level generation of the operator $\mathcal{O}_{qque}$, leading to $p \to \pi^0 e^+$ as a leading signature \cite{Dorsner:2012nq,Fajfer:2023gfi}. The expression for the matching estimate in this case arises from the 1-loop running from $\mathcal{O}_3 = \bar e^\dagger \bar u^\dagger \bar u^\dagger \bar d^\dagger$ onto $\mathcal{O}_{qque}$, given by
\begin{equation}
  C_{qque}^{pqrs} = \frac{1}{16\pi^2} [y_d]_{p^\prime} [y_u]_{q} 
  V^*_{p^\prime p} 
  C_{3}^{s [r q] p^\prime}\; \log \left(\frac{v}{M}\right) \; .
\end{equation}
According to our computations, the best limit for $\omega_4$ comes from $C_3^{1311}$, in agreement with the numerical result presented in Refs.~\cite{Gisbert:2024sjw,Fajfer:2023gfi} up to $\mathcal{O}(1)$ factors.

Similarly, the diquark scalar $\omega_2$ and the singly-charged dilepton scalar $\mathcal{S}_1$, couple antisymmetrically to pairs of fermions at the renormalisable level.\footnote{The colour sextet $\Omega_1$ also couples antisymmetrically to the $(QQ)$ bilinear. However, it additionally allows a flavour-unconstrained renormalisable coupling to $(\bar d \bar u)$. For each phenomenon of study, we employ a different renormalisable operator, and in neither case does the flavour antisymmetry impact the limit quoted.} As a consequence, the freedom to choose Yukawa matrix elements to obtain the most conservative limit is lost. This constraint does not alter the results in the case of $\mathcal{S}_1$, but it does for $\omega_2$ where the antisymmetry in flavour space forces us to involve the strange quark Yukawa coupling in the matching estimate. As a result, the upper bound derived from neutrino mass constraints is reduced by two orders of magnitude compared to the case without the flavour antisymmetry. Alternatively, nucleon decay limits remain unaffected by this antisymmetry, as the strange-violating decay channel $p \to K^{+} \nu$ exhibits comparable future sensitivity compared to the strange-conserving mode $p \to \pi^0 e^{+}$.

\subsection{Scalars with neutral components} \label{sec:scalar-veved}

When deriving limits on the UV scales $\Lambda_{\Delta L=2}$ and $\Lambda_{\Delta B=1}$, we also examine the possibility that the electrically neutral component of certain scalar multiplets acquires a vacuum expectation value (VEV). Specifically, we consider the scalar colour singlets $\Xi$, $\Xi_1$, $\Theta_1$, and $\Theta_3$.\footnote{For a recent analysis of Majorana neutrino masses induced by the electroweak quadruplets see Refs.~\cite{Giarnetti:2023dcr,Giarnetti:2023osf}.} However, we do not consider this scenario for the 2HDM field $\varphi$, as for any $\varphi$SMEFT operators to be genuine the antisymmetric contraction $\epsilon_{ij}H^i\varphi^j$ in $\mathrm{SU}(2)_{\rm{L}}$ is required to be present in the operator. This explicit contraction involves a charged component of either $H$ or $\varphi$, preventing the phenomena from occurring at tree level when both scalars acquire a VEV. While these effects could in principle arise at loop level~\cite{Zee:1980ai}, we dismiss the study of such a case and restrict ourselves to the tree level.

For the rest of the colour singlet scalars, we consider two distinct scenarios: (i) imposing a VEV by hand, which is constrained by electroweak precision tests (EWPTs) to obey $\langle X^0 \rangle <1$ GeV in order to not spoil \emph{custodial symmetry}~\cite{Veltman:1977kh,Sikivie:1980hm} or (ii) a VEV generated after Electroweak Spontaneous Symmetry Breaking (EWSSB), induced by the SM Higgs doublet VEV through the scalar potential via linear terms in the new fields. We focus on the latter case, as it offers greater flexibility in tuning the parameters of the theory while accommodating a viable range of upper bounds on the UV scales $\Lambda$.

Take for instance the scalar $\Xi$, whose renormalisable (trilinear) coupling is given by $ \mu \;\Xi H^\dagger H$. After EWSSB, this coupling along with the mass term for $\Xi$ will induce a VEV for $\Xi$ of the kind $\langle \Xi^0 \rangle \sim \mu \; v^2/M^2$,\footnote{For the SU(2)$_{\rm{L}}$ quadruplets $\Theta_{1,3}$, the induced VEV had the analytical expression $\langle \Theta_{1,3}^0 \rangle \sim \lambda \; v^3/M^2$, but analogous conclusions apply.\label{fn:inducedv}} for which EWPTs demand $\langle \Xi^0 \rangle \lesssim \mathcal{O}(1)$ GeV. In this setting, the same VEV enters in both $\Delta L=2$ and $\Delta B=1$ operators, so when compared to the experimental constraints on the absolute scale $m_\nu$ and assuming a future signal at Hyper-K, the question of how feasible it is to keep $\mathcal{O}(1)$ WCs $c$ is interesting to address. In this particular case, building on the premises of this paper, where we assume that the main contribution to $m_\nu$ is given by the operators listed in the tables of App.~\ref{sec:appendixtables}, we can write the expressions for $\Lambda_{\Delta L=2}$ and $\Lambda_{\Delta B=1}$ in terms of the parameters $\mu$ and $M$, which are linked through the constraint from EWPTs, such that $\Lambda$ is bounded from above as:
\begin{enumerate}
    \item $\displaystyle \Lambda_{\Delta L=2} \lesssim 5 \cdot 10^{4} \, c_{\Delta L=2}^{1/2} 
    \left( \frac{\langle \Xi^0 \rangle}{1 \; \rm GeV}\right)^{1/2} \left(\frac{0.05 \; \rm eV}{\sqrt{\Delta m_{\rm atm}^2}}\right)^{1/2}$ TeV,
    \item $\displaystyle \Lambda_{\Delta B=1} \lesssim 10^5 \, c_{\Delta B=1}^{1/4}
    \left( \frac{\langle \Xi^0 \rangle}{1 \; \rm GeV} \right)^{1/4}
    \left( \frac{\tau_p}{10^{35} \; \rm years} \right)^{1/8}
     \;$ TeV,
\end{enumerate}
where we have used the nuclear matrix element $\alpha \sim 0.01$ GeV$^3$~\cite{Yoo:2021gql} and defined $\langle \Xi^0\rangle = \mu \,v^2/M^2   $. In this scenario, it would require a tuning of the order of $c_{\Delta B=1}/c_{\Delta L=2} \sim 10^{-2}$ for $\mathcal{O}(1)$ $c_{\Delta L=2}$ in order to obtain $\Lambda_{\Delta B=1} \sim \Lambda_{\Delta L=2}$. We may also go to the limiting case where $M \sim \mu \sim \Lambda$ and EWPTs are saturated, giving
\begin{enumerate}
    \item $\displaystyle \Lambda_{\Delta L=2} \sim 5 \cdot 10^3 \; c_{\Delta L=2}^{1/2}$ TeV,
    \item $\displaystyle \Lambda_{\Delta B=1} \sim 10^5 \; c_{\Delta B=1}^{1/4}$ TeV .
\end{enumerate}
Note that, as expected since it is a general relation, the previous tuning for $c_{\Delta B=1}/c_{\Delta L=2}$ still applies, and we can check that for $\mathcal{O}(1)$ $c_{\Delta L=2}$ EWPTs are satisfied.

The same line of reasoning applies to the rest of the multiplets, modulo the analytic expressions for the induced VEV as mentioned in footnote~\ref{fn:inducedv}. In general, when we are in the limiting case, we can also set an absolute lower bound on the scales $\Lambda$ from EWPTs saturations for $c \sim \mathcal{O}(1)$. For triplets (quadruplets) under SU(2)$_{\rm L}$, this absolute bound corresponds to $\Lambda \gtrsim 30 $ $(2)$ TeV. If we are not in the limiting case, we can always keep $\mu \ll M$ in agreement with EWPTs.

Therefore, in the tables of App.~\ref{sec:appendixtables} we present only the limiting case where we set $\mu \sim M \sim \Lambda$, consistent with our approach for all other results. Additionally, for this case, we also provide the value of the EWSSB induced VEV. For completeness, we quote here the upper bounds on $\Lambda_{\Delta L=2}$ ($\Lambda_{\Delta B=1}$) from EWPTs, which are $5 \cdot 10^{12} \;(10^7)$, $ 5 \cdot 10^4\;(10^5)$, $ 5 \cdot 10^9 \;(10^7)$, and $5 \cdot 10^2 \;(10^7)$ TeV for the scalars $\Xi_1, \; \Xi, \; \Theta_3$, and $\Theta_1$, respectively.

\section{Results}\label{sec:results}

Below we present the main results of our study and their analysis. We adopt a compact notation for the nature of the mechanisms of $B$ and $L$ violation throughout the section: $\mathbf{T}_{i}$ and $\mathbf{L}_{i}$ stand for the process $i \in \{\Delta L =2,\Delta(B-L)=0, \Delta(B-L)=2\}$ generated at tree level or loop level, respectively. The tables of App.~\ref{sec:appendixtables} contain many of the details of our results and we refer to them throughout the section where necessary.

\subsection{Upper bounds}

Table~\ref{tab:generalclassification} summarises the possible mechanisms by which each LSME generates Majorana neutrino masses and induces nucleon decay. When a given multiplet mediates a $\Delta L=2$ or $\Delta (B-L)=0$ process at tree level, $M$ can be directly associated with the scale at which BNV or LNV occurs in the UV. However, if the multiplet induces $\Delta(B-L)=2$ nucleon decay, an additional UV scale, $\Lambda_{\rm{\Delta B=1}}$, is required. Consequently, we distinguish between $\Delta (B-L)=0, 2$. Notably, for the category (\textbf{T$_{\Delta L =2}$}, \textbf{T$_{\Delta(B-L)=2}$}) only two particles, the type-I and type-III seesaw mediators $N, \; \Sigma$, might mediate $\Delta(B-L)=2$ nucleon decay at tree-level~\cite{Li:2022abx,Li:2023cwy}. Importantly, no single particle can simultaneously mediate two-body nucleon decay via dimension-6 SMEFT operators while also generating Majorana neutrino masses at tree level. One easily notices that while only three seesaw particles $N$, $\Sigma$, and $\Xi_1$ generate neutrino masses at tree level, a significantly larger number contribute to tree-level nucleon decay. Specifically, 4 (13) particles induce BNV nucleon decay at tree level via dimension-6 (dimension-7) operators, as found in Refs.~\cite{deBlas:2017xtg,Li:2023cwy,Li:2022abx} and summarised in Tab.~\ref{tab:generalclassification}. Among the others, nine fall into the (\textbf{L$_{\Delta L =2}$, L$_{\Delta (B-L) = 0}$}) category, while the remainder belong to (\textbf{L$_{\Delta L =2}$, L$_{\Delta (B-L) = 2}$}).

{
    \setlength{\arrayrulewidth}{0.2mm}
    \setlength{\tabcolsep}{5pt}
    \renewcommand{\arraystretch}{1.5}
    
    \begin{table}[ht]
    \centering
    \large
    \begin{tabular}{|c|c|c|}

       \hline & \textbf{T$_{\Delta L=2} $ } & \textbf{L$_{\Delta L=2}$} \\ \hline
       
       \textbf{T$_{\Delta(B-L)=0}$}
       & None
       & $\omega_1, \; \zeta, \; \mathcal{Q}_1, \; \mathcal{Q}_5$ \\ \hline

       \multirow{2}{*}{\textbf{T$_{\Delta(B-L)=2}$}}
       & \multirow{2}{*}{$ N, \; \Sigma$} &  $\Pi_1, \; \Pi_7, \; \omega_2, \; \mathcal{U}_2, \; \mathcal{X}$\\
       & & $T_1, \; Q_1, \; Q_5, \;D, \; E, \; \Delta_1$ \\ \hline
    
       \multirow{2}{*}{\textbf{L$_{\Delta (B-L) = 0}$}}
       & \multirow{2}{*}{$\Xi_1$} & $\omega_4 \;, \varphi \;, \Omega_4, \;U,\; Q_7$ \\
       & & $\Delta_3 ,\; T_2 \;, \Sigma_1,\; \mathcal{L}_1$ \\ \hline
       
       \multirow{3}{*}{\textbf{L$_{\Delta (B-L)=2}$}}
       & \multirow{3}{*}{None}  &
       $\Upsilon$, $\Phi$, $\Omega_2$, $\Omega_1$, $\mathcal{S}_1$, $\mathcal{S}_2$ \\ 
       & & $\Xi$, $\Theta_3$, $\Theta_1$, $\mathcal{Y}_1$, $\mathcal{Y}_5$, $\mathcal{G}_1$, $\mathcal{G}$\\
       & & $\mathcal{B}$, $\mathcal{B}_1$, $\mathcal{H}$, $\mathcal{W}$, $\mathcal{W}_1$, $\mathcal{L}_3$, $\mathcal{U}_5$ \\\hline

       \end{tabular}
       
       \caption{General classification of the particles based on their role in generating both Majorana neutrino masses and nucleon decay: \textbf{T}$_i$ and \textbf{L}$_i$ stands for the process $i$ generated at tree level or loop level, respectively.
       }
    \label{tab:generalclassification}
    \end{table}
}

\begin{figure*}[ht]
    \centering
\includegraphics[width=0.72\textwidth]{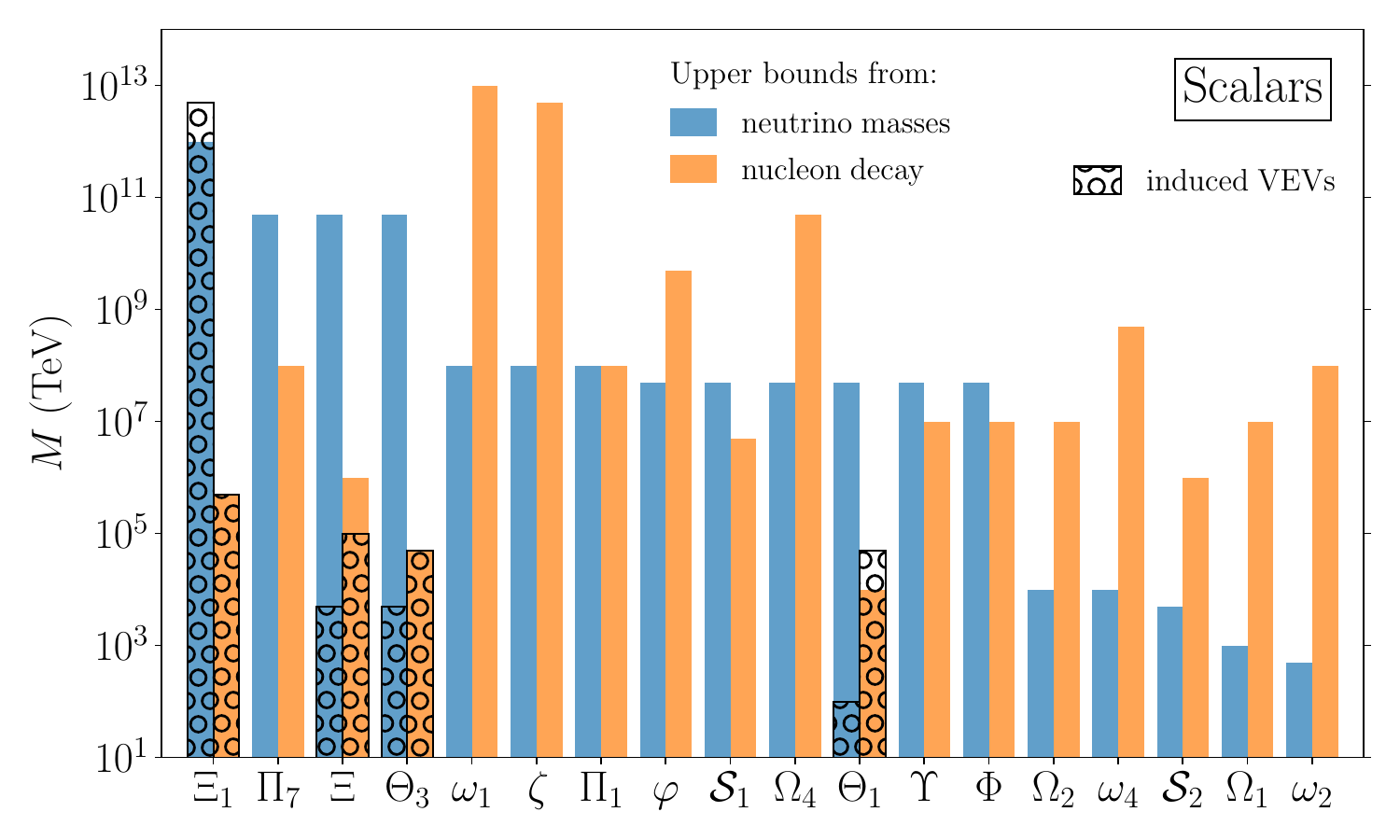}\\
    \vspace{0.1cm}    \includegraphics[width=0.72\textwidth]{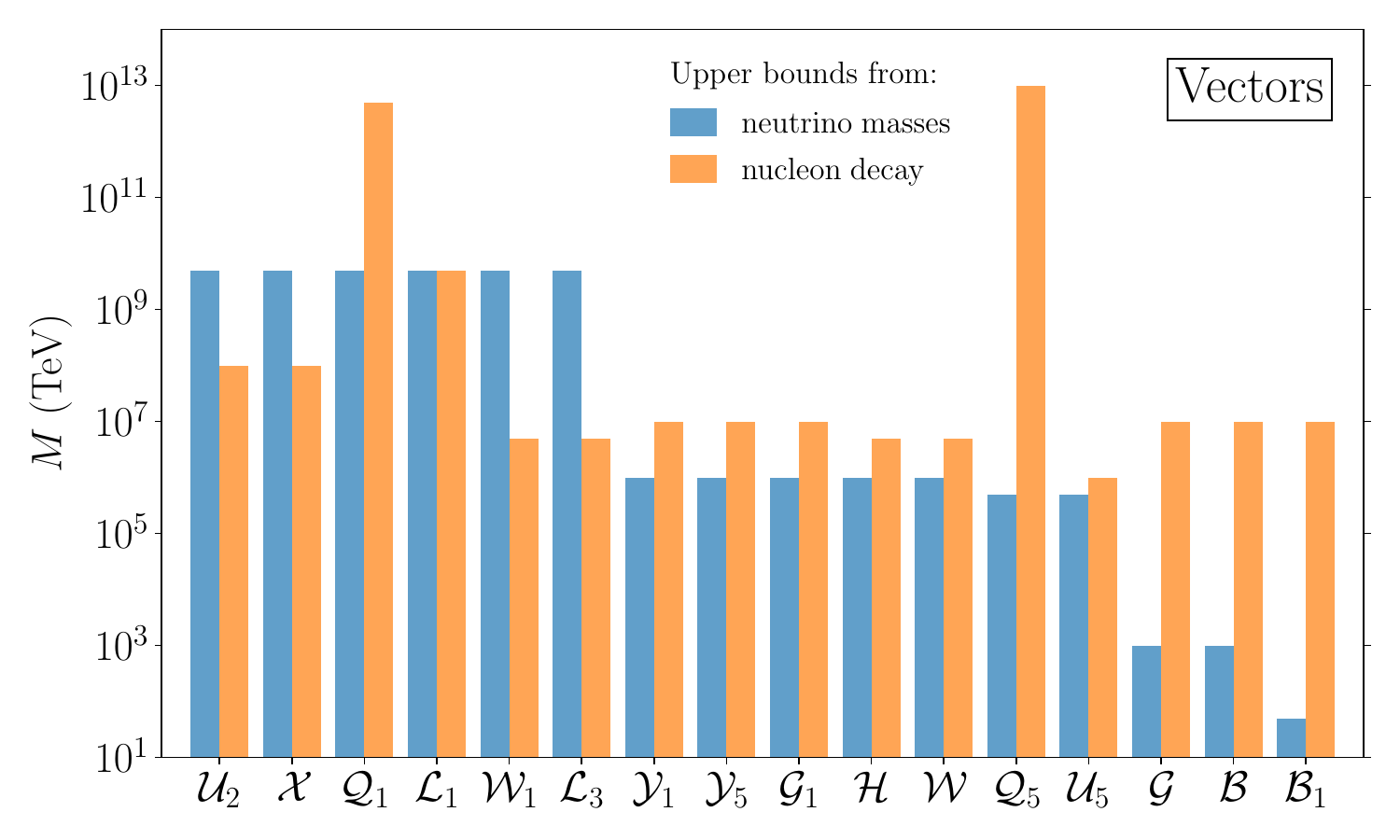}\\
    \vspace{0.1cm}
\includegraphics[width=0.72\textwidth]{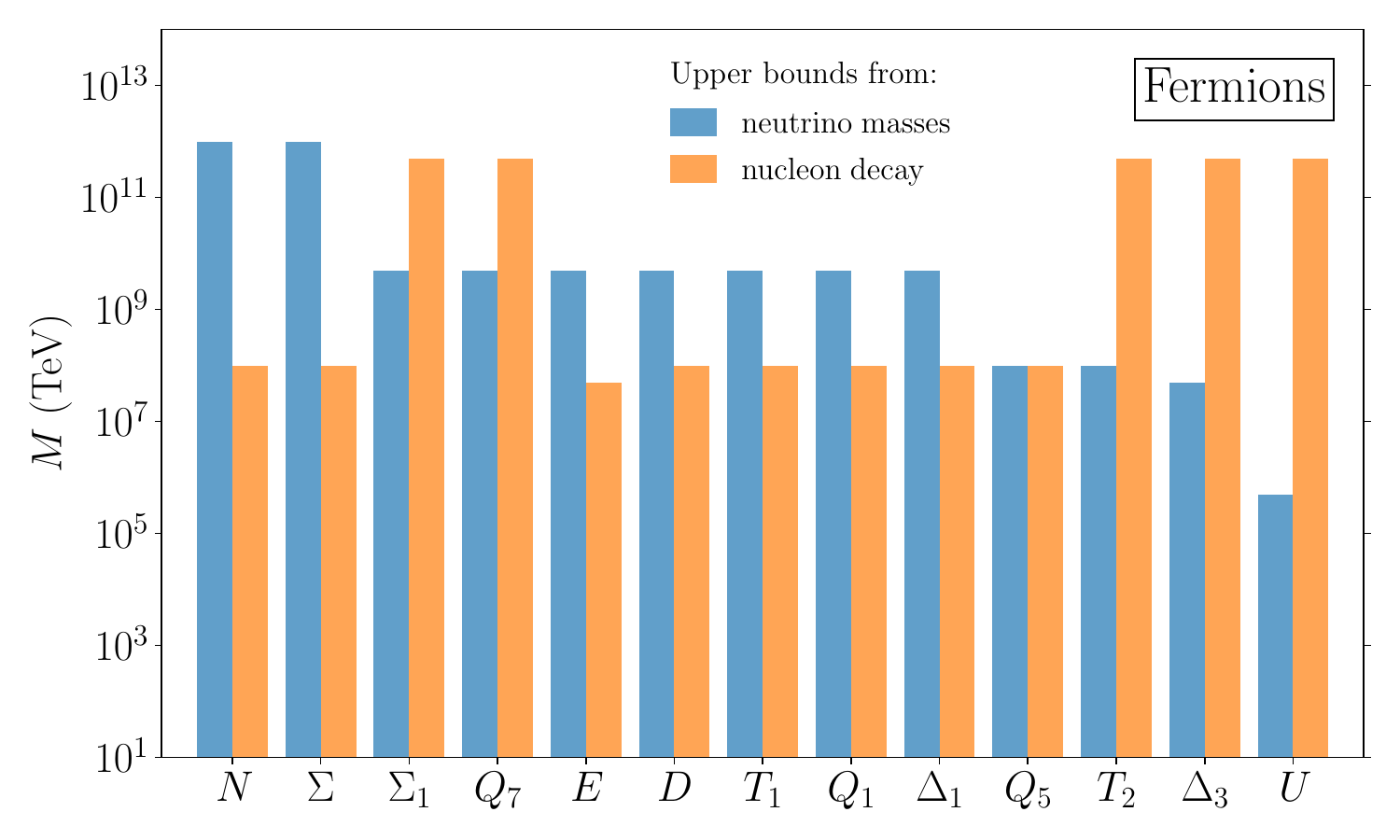}
    \caption{Upper limits on the mass of new scalars (top), vectors (middle) and fermions (bottom) that couple linearly to the SM, obtained from reproducing neutrino masses (in blue) and a hypothetical signal from nucleon decay (in orange).}\label{fig:barplots}
\end{figure*}

In Fig.~\ref{fig:barplots} we present the upper bounds on the masses of the 47 LSMEs under study, organised into three subfigures according to their transformations under the Lorentz group. As explained in Sec.~\ref{sec:genuineness}, we exclude the singlet $\mathcal{S}$ from our analysis. For each LSME, we display the limits from nucleon decay (orange) and Majorana neutrino masses (blue). Additionally, for scalar LSMEs that may acquire a VEV, we indicate the corresponding upper bound with a bubble region, assuming this occurs. The details of the bound computation, including the loop order at which the process is generated for each LSME and the explicit suppression factors, are provided in App.~\ref{sec:appendixtables}.

\begin{figure*}[t]
    \centering
    \includegraphics[width= 0.9\columnwidth]{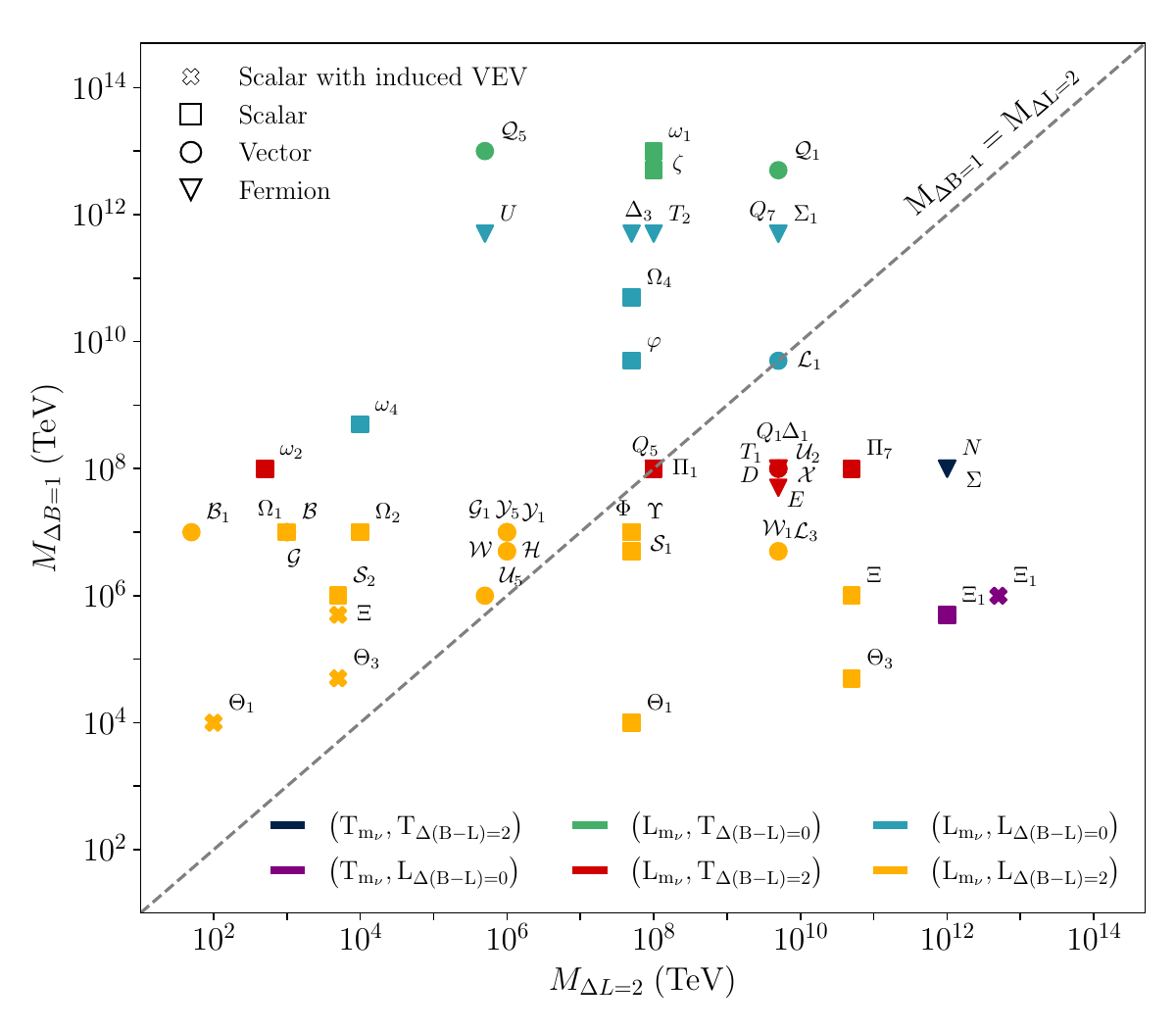}
    \caption{Two-dimensional plane $(M_{\Delta L=2},$ $M_{\Delta B=1})$ obtained under the assumption $M\sim \Lambda$. We use different colours to denote each LSME according to their category in Tab.~\ref{tab:generalclassification}, and we use different markers to denote their Lorentz transformation. Note that several LSMEs share common points in the plot. See the main text for further details.
    \label{fig:parameterspace}
    }
\end{figure*}

In Fig.~\ref{fig:parameterspace} we present the limits for each particle in the two-dimensional plane $(M_{\Delta L=2},$ $M_{\Delta B=1})$. Different colours are used to classify LSMEs according to the categories of Tab.~\ref{tab:generalclassification}. We observe that the LSMEs in the upper (rightmost) region in green (black and purple), correspond to the LSMEs that generate nucleon decay (Majorana neutrino masses) at tree level via dimension-6 (dimension-5) SMEFT operators, and any other LSME has a lower mass limit, as shown in the Fig.~\ref{fig:barplots}.

As summarised in Tab.~\ref{tab:generalclassification}, out of the 47 LSMEs, the majority induce nucleon decay at most at the loop level, with nearly all doing so at just one loop. Only six scalar LSMEs, \textit{i.e.} $\Omega_4$, $\Xi_1$, $\varphi$, $\Xi$, $\Theta_1$, and $\Theta_3$, generate nucleon decay at a minimum of two loops, leading to a more suppressed limit on their masses. However, these conclusions do not directly extend to the generation of Majorana neutrino masses, as many particles contribute to them at two loops or even three loops~\cite{Herrero-Garcia:2019czj}. Of particular interest are the four LSMEs $\Omega_1$, $\mathcal{B}$, $\mathcal{B}_1$, and $\mathcal{G}$, that generate this phenomenon at least at three loops.
    
In Fig.~\ref{fig:parameterspace} one may observe a general pattern in the generation of both Majorana neutrino masses and nucleon decay. The LSMEs are typically ordered from right to left based on the loop order at which they generate Majorana neutrino masses, with the leftmost region corresponding to LSMEs that generate them at three loops, followed by those that do so at two loops, and so forth. However, this ordering (which would now be from top to bottom) does not hold for nucleon decay, as each LSME matches onto BNV operators of different mass dimension in the SMEFT, namely, dimension-6 and dimension-7, each suppressed by either two or three powers of $\Lambda$. As a result, two distinct regions emerge: one around $10^{12}$ TeV for LSMEs matching onto dimension-6 operators and another around $10^7$ TeV for those matching onto dimension-7 operators. Nonetheless, a pattern can be identified for particles belonging to the \textbf{L$_{\Delta (B-L) = 0}$} category, which, due to the suppression factors, always appear between \textbf{T$_{\Delta (B-L) = 0}$} (in green) and \textbf{T$_{\Delta (B-L) = 2}$} (in red) LSMEs. This follows from selecting the $X$SMEFT operator with the leading contribution at a given dimension, \textit{i.e.} the least suppressed by loop, Yukawa, CKM, and $W$ boson exchanges factors. A particularly notable case is $\Xi_1$, which exhibits significant suppression due to two down-type quark Yukawa insertions, the associated CKM matrix elements, and the antisymmetric structure in flavour space for right-handed down-type quarks in the non-renormalisable operator. Interestingly, a higher-dimensional $\Xi_1$SMEFT operator could yield a slightly stronger limit. However, since our primary focus is on \emph{simple} UV models, we restrict our analysis to the lowest-dimensional operator.

\subsection{Combining nucleon decay and neutrino masses}

In this section, we examine the special case in which all relevant scales align. In this approximation, one can assess the degree of fine-tuning required for the dimensionless Wilson coefficients associated with either $\Delta L=2$ and $\Delta B = 1$ operators to account for all observed phenomena within the same UV framework. This provides insight into the viability of embedding both mechanisms within a unified UV theory, such as a GUT. This information can be read from the dashed diagonal line in Fig.~\ref{fig:parameterspace}, which may be useful for identifying those LSMEs where the limits from both phenomena are comparable. LSMEs lying close to the diagonal line could potentially allow for simple UV completions that explain both Majorana neutrino masses and nucleon decay with $\mathcal{O}(1)$ dimensionless couplings in the limiting case $M \sim \Lambda$. A detailed exploration of this scenario is beyond the scope of this work. 
    
\begin{figure}[th!]
    \centering
    \includegraphics[width=\columnwidth]{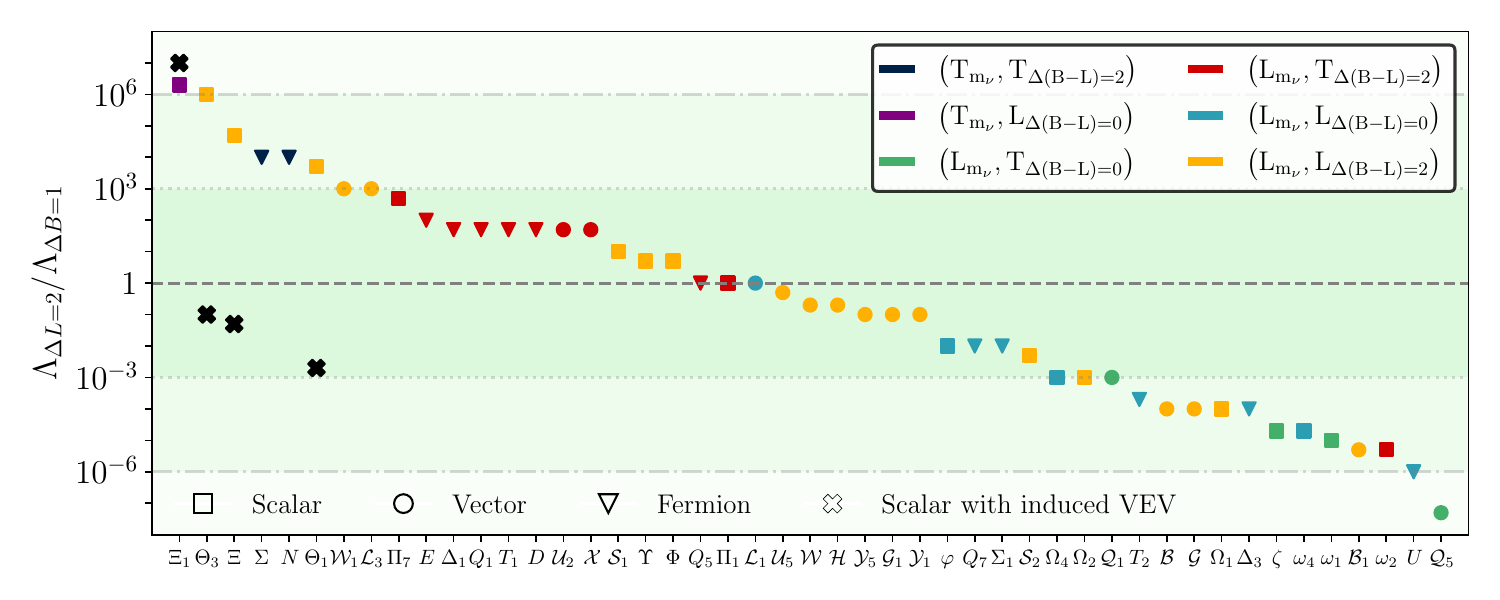}
    \caption{Ratio of $\Lambda_{\Delta L=2}$ and $\Lambda_{\Delta B=1}$ for each LSME. For readiness, we highlight different ratios of UV scales with green bands. See main text for details.}
    \label{fig:lambdasdivision}
\end{figure}

In Fig.~\ref{fig:lambdasdivision}, we illustrate this limiting case comparing the limit obtained for $\Lambda_{\Delta B=1}$ and $\Lambda_{\Delta L=2}$ for each LSME. In particular, those for which the ratio of these two UV scales is closer to unity are more naturally accommodated within a single UV theory capable of explaining both Majorana neutrino masses and baryon-number-violating processes while keeping $\mathcal{O}(1)$ couplings. We can use $\Pi_1$ to exemplify this result. This BSM multiplet falls into the category (\textbf{L$_{\Delta L =2}$, T$_{\Delta (B-L) = 2}$}), leading to the presence of three distinct UV scales in our analysis: $M, \; \Lambda_{ \Delta B=1}$, and $\Lambda_{\Delta L=2}$. Assuming this particle dominantly contributes to nucleon decay and neutrino masses through the operators listed in Tabs.~\ref{tab:fermion-neutrino-masses}--\ref{tab:vector-proton-decay} and imposing both the atmospheric neutrino mass scale and a potential future signal at Hyper-K, it is possible to account for both phenomena by setting a single UV scale $M \simeq \Lambda_{\Delta B=1} \simeq \Lambda_{\Delta L=2} \simeq 10^{8}$ TeV while keeping $\mathcal{O}(1)$ dimensionless WCs $c$ and $y$. If instead, we keep $M \leq \Lambda_{\Delta B=1} = \Lambda_{\Delta L=2}$, both BSM effects can still be explained, provided that

\begin{equation}
       \frac{c_{\Delta \rm{B} = 1}}{c_{\Delta \rm{L} = 2}} \lesssim \left( \frac{M}{10^8 \; \rm{TeV}}\right)^2. \label{eqRelMassPi1}
\end{equation}

The same conclusions apply to all particles close to the horizontal line in Fig.~\ref{fig:lambdasdivision}. For those particles in the left (right) region, namely $\Lambda_{\Delta L=2} \gg \; \Lambda_{\Delta B=1}$ ($\Lambda_{\Delta L=2} \ll\; \Lambda_{\Delta B=1}$), we require a large suppression in $c_{\Delta L=2}$ ($c_{\Delta B=1}$) compared to $c_{\Delta B=1}$ ($c_{\Delta L=2}$). Consequently, particles for which the ratio of UV scales is close to unity can \emph{naturally} accommodate both phenomena for a unique UV scale.\footnote{This statement relies on the choice of the operators in Tabs.~\ref{tab:fermion-neutrino-masses}--\ref{tab:vector-proton-decay}. However, if we use our assumptions and list the simplest $X$SMEFT operator, this conclusion holds. Note that $X$SMEFT operators of the same mass dimension could mediate both phenomena, but less dominantly, since we list the dominant contribution, relaxing the conclusions stated above.} Analysing all particles in detail would go against the logic of this work, where simplicity and (relatively) model independence are used as premises. Therefore, we encourage the reader to interpret Fig.~\ref{fig:lambdasdivision} with caution.

A similar line of reasoning applies to particles that generate one of the processes at tree-level on their own, namely those three in the \textbf{T$_{\Delta L =2}$} category and the four in \textbf{T$_{\Delta (B-L) = 0}$}. For the three (seesaw) particles in \textbf{T$_{\Delta L =2}$} category, if we assume $\mathcal{O}(1)$ WCs, the atmospheric neutrino mass scale imposes a (conservative) upper bound on the scale $M \lesssim 10^{12}$ TeV. Consequently, for these particles to be responsible for a potential future signal of nucleon decay in Hyper-K, we would require $c_{\Delta B = 1} \gg c_{\Delta L = 2}$, which would break perturbativity, as $c_{\Delta L = 2}$ is assumed to be $\mathcal{O}(1)$. Similarly, for the four particles in the \textbf{T$_{\Delta (B-L) = 0}$} category, if we assume $\mathcal{O}(1)$ WCs and a positive future signal in Hyper-K, we obtain a mass bound of approximately $10^{13}$ TeV for the (leptoquark) particle. Therefore, for these particles to be the dominant source of Majorana neutrino mass generation, we would need $c_{\Delta L = 2} \gg c_{\Delta B = 1}$, which would again break perturbativity in the same way as before. Thus, we conclude that for these particles to be able to generate both phenomena, their mass should be below the lowest upper bound of the two (\textit{i.e.} for instance that corresponding to $m_\nu$), with a suppression in the WC of the other (\textit{i.e.} the proton decay one in this case).

A different scenario arises for particles that generate one or both of these processes at loop level. In such cases, the presence of three distinct energy scales allows for the accommodation of both phenomena within the perturbative regime. In general, the interplay between these effects leads to a relation between the UV scales $\Lambda_{\Delta B=1}$ and $\Lambda_{\Delta L=2}$ analogous to Eq.~\eqref{eqRelMassPi1}. Consider, for example, $\omega_2$ in the (\textbf{L$_{\Delta L =2}$},\textbf{T$_{\Delta (B-L) = 2}$}) category. This particle alone does not induce nucleon decay, as its renormalisable interaction with the SM content is not sufficient, and a dimension-5 $\omega_2$SMEFT operator is needed to trigger $\Delta B=1$ processes, as displayed in Tab.~\ref{tab:scalar-proton-decay}. Thus, generating nucleon decay at tree level via the dimension-7 SMEFT operator $\mathcal{O}_{\bar l dud \tilde H}$ requires an additional particle, such as $\Pi_7$, $Q_1$ or $N$~\cite{Li:2022abx,Li:2023cwy}, which are assumed to be heavier than $\omega_2$ by hypothesis. Assuming $\mathcal{O}(1)$ WCs and imposing future sensitivities for the dominant nucleon decay channel $p \to K^+ \nu$ generated at tree-level from this particle, we find a relation of the kind
\begin{equation}
    \frac{M^2_{\omega_2} \cdot \Lambda_{\Delta B=1}}{(10^8 \;\rm{TeV})^3} \cdot \left( \frac{10^{35} \; \rm years}{\tau_p}\right)^{1/2} \lesssim 1,\label{eqBoundomega2}
\end{equation}
where we have used the nuclear matrix element $\alpha \sim 0.01$ GeV$^3$~\cite{Yoo:2021gql}. Similarly, assuming $\mathcal{O}(1)$ WCs and imposing the atmospheric neutrino mass scale we find\footnote{Notice how, in this case, the antisymmetry in the flavour of the renormalizable term for $\omega_2$ translates into different down-type Yukawas appearing in the loop-matching contribution, as shown in Tab.~\ref{tab:scalar-neutrino-masses}, leading to a more stringent upper bound, as detailed in Sec.~\ref{sec:flavoursymmetries}.}
\begin{equation}
    \frac{\sqrt{\Delta m^2_{\rm{atm}}}}{0.05 \; \rm eV} \cdot \frac{\Lambda_{\Delta L=2}}{500 \; \rm TeV} \lesssim 1
\end{equation}
By requiring $M_{\omega_2} \leq \Lambda_{\Delta L=2}$, this translates into an upper bound of the kind $M_{\omega_2} \leq 100$ TeV. This bound on $M_{\omega_2}$, when saturated, satisfies Eq.~\eqref{eqBoundomega2}.

We emphasise once again the highly conservative nature of the upper limits shown in Fig.~\ref{fig:barplots}, which arise from several factors, including:
\begin{enumerate}
    \item The dimensionless WCs $y$, and $c$ are universally set to unity and assumed to be flavour-independent. However, most of these coefficients, particularly those appearing on the right side of the third column in Tabs.~\ref{tab:fermion-neutrino-masses}--\ref{tab:vector-proton-decay}, originate from the product of multiple Yukawa-like couplings to SM fermions. In the SM, such couplings are typically much smaller than unity, except for $y_t$. Smaller values for these dimensionless couplings would result in stronger bounds on the UV scale, as expected. This effect may be especially relevant for $\Delta B=1$ processes, where all involved quarks are light.
    \item To derive bounds, we set $M \sim \Lambda$, which further pushes our limits into the UV regime, specifically toward the saturation bound under our assumption that $M \leq \Lambda$. A clear example of the impact of this assumption arises in the case of particles in the category (\textbf{L$_{\Delta L =2}$, T$_{\Delta (B-L) = 2}$}). Notably, in any UV-complete model that generates $\Delta (B-L) = 2$ proton decay at tree level, at least two LSMEs are required: either two scalars, a vector-like fermion and a scalar, or a vector-like fermion and a vector. In such scenarios, nucleon decay experiments constrain the product of masses, such as $M_F \cdot M_{S,V}^2 $ or $M_{S_1}^2 \cdot M_{S_2}^2/\mu $. As a result, the lightest LSME, which we consider throughout this work, can have a mass of a few TeV without conflicting with other observables, particularly those related to flavour physics, while the heavier multiplet can live at a much higher UV scale, approaching $10^{12}$ TeV.
\end{enumerate}
Under the previous assumptions, it is possible to list the combination of operators that lead to the dominant contribution to such phenomena, but if, for any reason any of the dimensionless WCs is suppressed, or there are large scale separations between new UV physics, the dominant contribution could arise from other combinations of operators. However, in such cases, the upper bound would be smaller than the one quoted in this work.

For completeness, let us also comment on the qualitative different behaviour arising from the distinct nucleon decay phenomenology, \textit{i.e.} $\Delta (B-L)= 0, \,2$. The low-energy operators leading to those phenomena are either dimension 6 or 7, thus suppressed by two or three powers of a UV energy scale, respectively. As illustrated in Fig.~\ref{fig:parameterspace}, regardless of the mass dimension of the $X$SMEFT operator, the corresponding limits cluster into two regions: one around $10^7$ TeV (corresponding to those LSMEs that induce $\Delta (B-L)=2$ nucleon decay) and another around $10^{12}$ TeV (corresponding to those LSMEs that induce $\Delta (B-L)=0$ nucleon decay). Therefore, those LSMEs leading to $\Delta (B-L)=2$ nucleon decay are further suppressed compared to those of $\Delta (B-L)=0$. It might be the case that a UV theory contains one of the LSMEs that generates $\Delta (B-L)=2$ nucleon decay (either at tree or loop level), but does not contain the rest of particles necessary to induce it. In such a case, in order for this LSME to be responsible for proton decay, one would need to go to a higher dimension in $X$SMEFT, effectively leading to a combination of $X$SMEFT operators matching at loop-level onto a dimension-6 $\Delta B=1$ SMEFT operator. In this case, it might be the case that the bounds on $\Lambda_{\Delta B=1}$ quoted in this work for such LSMEs are not the most conservative ones, but let us remind that we keep ourselves to the lowest-dimensional $X$SMEFT operators, motivated by \emph{simplicity}, and going into higher dimension will in general imply more intricate UV theories. An example of such a case is $T_1$, which induces $\Delta (B-L)=2$ nucleon decay through the combination of $y_p \bar T_1 Q_p H^\dagger $ + $ c_{qrs} (T_1 Q_q) (L^\dagger_r \bar d^\dagger_s)$. From this, we obtain an upper bound on its mass of $10^8$ TeV. However, if the UV theory does not include the LSMEs $\mathcal{X}$ or $\Pi_1$,\footnote{Note that $\mathcal{Q}_5$ also UV completes the non-renormalisable operator but would lead to nucleon decay through a dimension-6 SMEFT operator at tree level.} one must consider higher-dimensional operators, matching at the loop level onto a $\Delta (B-L)=0$ dimension-6 operator. Consequently, this leads to a more conservative upper bound on the mass of $T_1$, given by $5 \cdot 10^{11}$ TeV.

Finally, we note that all LSMEs analysed in this work can be embedded within minimal GUTs and supersymmetric scenarios, such as SU(5) where $\omega_1, \phi$ are contained in the $\textbf{5}_H$ representation, $\mathcal{Q}_5$ in the $\textbf{24}_V$ representation, and $\Phi, \Omega_1, \zeta, \Pi_7, \omega_4$ in the $\textbf{45}_H$ representation, to name a few examples.\footnote{We refer the reader to Ref.~\cite{Hati:2018cqp} and Tab.~1 of Ref.~\cite{Gargalionis:2024jaw} for particular GUT embeddings and known models containing LSMEs considered here.} Similarly, certain particles appear in well-established radiative neutrino mass generation frameworks, such as $\mathcal{S}_1$ and $\mathcal{S}_2$ in the Zee--Babu model \cite{Babu:1988ki,Zee:1980ai}. We expect that the combined bounds derived in this work will serve as a guide for constraining UV-complete models incorporating these new multiplets.

\section{Conclusions}\label{sec:conclusions}

Particles beyond the SM often violate its accidental symmetries, particularly baryon and/or lepton number. In this work, we explored scenarios where these low-energy symmetries are broken at a high scale $\Lambda$, involving at least one multiplet with a mass $M$ that satisfies $v < M < \Lambda$. The particles considered in this study couple linearly to the SM field content at the renormalisable level. In addition to these couplings, we also consider effective operators that, in combination with renormalisable interactions, give rise to Majorana neutrino masses and BNV nucleon decay. Out of all the possible effective operators of such kind, we are interested in the \emph{simplest} and \emph{genuine} operators linear in such multiplets. These operators represent the most minimal classes of models that generate neutrino masses (via the dimension-5 Weinberg operator) and induce nucleon decay (through baryon-number-violating operators up to dimension 7). 

Under the assumption of a proton decay signal in one of the upcoming experiments, we derived an upper bound on the intermediate scale $M$ for each field 
using this framework, and under the assumption that these multiplets provide the dominant contributions to neutrino masses and hypothetical nucleon decay signals. Our results show that, whereas the particles may have $\Delta L=2$ scales that span all possible energy scales, from hundreds of TeVs to GUT scales, the BNV energy scales typically cluster around either $10^7$ TeV or $10^{12}$ TeV. This can be traced back to the symmetry properties of the UV theory. UV theories with $\Delta (B-L)=0$ generate the $\Delta (B-L)=0$ dimension-6 SMEFT operators, which result in upper limits of order $10^{12}$ TeV, while UV theories with $\Delta (B-L)=2$ generate the $\Delta(B-L)=2$ dimension-7 SMEFT operators, which result in upper limits of order $10^7$ TeV.
Using the framework described in this work, we classify these multiplets based on the derived upper bounds on their masses. This classification highlights heavy multiplets that, if responsible for BNV nucleon decay and/or neutrino masses through the mechanisms studied here, could be probed in complementary searches. In particular, the conservative upper bounds of $\mathcal{B}_1$, $\omega_2$, $\Omega_1$, $\mathcal{B}$, and $\mathcal{G}$ derived from their generation of radiative neutrino masses are in the few-hundreds of TeV range, and therefore may be tested at future colliders. 

Our approach also allows us to determine the relative strengths of $\Delta L=2$ and $\Delta B=1$ processes in the UV, by assessing the ratio $\Lambda_{\Delta L=2}/\Lambda_{\Delta B=1}$, offering a systematic method to organise the landscape of UV models that violate these symmetries. Moreover, our results can serve as a valuable tool for model builders aiming to design UV theories that can explain both phenomena. Similarly, our analysis may also be used to classify possible new particles that couple to SM fields linearly and generate gauge coupling unification, in the line of Ref.~\cite{Hagedorn:2016dze}. 

Should a proton decay signal be observed in a specific channel at Hyper-K or DUNE, our study could play a pivotal role in guiding the development of simplified models capable of reproducing the signal, some of which may also generate neutrino masses.

\acknowledgments

We are deeply grateful to Arcadi Santamaria for insightful discussions throughout various stages of the project. We also thank Elena Bermejo for her contributions during the initial stages of this work. MS acknowledges support by the Australian Research Council through the ARC Discovery Project DP200101470.
JHG, JG, and ABB are partially funded by the Spanish ``Agencia Estatal de Investigación'', MICIN/AEI/10.13039/501100011033, through the grants PID2020-113334GB-I00, and in addition, JHG by PID2020-113644GB-I00 and by the ``Consolidación Investigadora'' Grant CNS2022-135592 and JG by the ``Juan de la Cierva'' programme reference FJC2021-048111-I, both funded also by ``European Union NextGenerationEU/PRTR''. This research is also financed by the ``Generalitat Valenciana'': JHG is also supported through the Plan GenT Excellence Program (CIESGT2024/7), while ABB is funded by grant CIACIF/2021/061. JG is also supported by the ARC Centre of Excellence for Dark Matter Particle Physics CE20010000. Feynman diagrams were
generated using the Ti\textit{k}Z-Feynman package for
\LaTeX~\cite{Ellis:2016jkw}.

\appendix

\section{Matching and running estimates} \label{sec:matching-and-running-estimates}

In this appendix, we aim to give the reader a more solid intuition for the ways in which the Weinberg operator or the $d \leq 7$ $B$-violating operators may arise at the low scale, as discussed in Sec.~\ref{sec:framework}. This includes a justification for why matching (and not running) dominates when the mass-dimension of the operator in the $X$SMEFT differs from that of the corresponding SMEFT operator by more than one unit, \textit{i.e.} when $\Delta d > 1$. Additionally, we motivate the forms of Eqs.~\eqref{eq:deltad-0} --~\eqref{eq:deltad-1} which distinguish different ways in which the running contribution can dominate at low scales. Here we specialise to the case of $\Delta L =2$ operators matching onto and running into the Weinberg operator for the sake of simplicity. We note that similar conclusions apply equally well, \textit{mutatis mutandis}, to the $d \leq 7$ $B$-violating operators considered in our study.

First, we highlight that the running contributions to the dimension-5 Weinberg operator defined in Eq.~\eqref{eq:weinbergop} $C_5$ are fixed by dimensional analysis to be\footnote{Recall that in dimensional regularisation, only infrared scales can appear in the numerator.}
\begin{equation} \label{eq:b1}
    C_{5,\mathrm{EFT}} \sim \frac{1}{\Lambda} \left(\frac{M}{\Lambda}\right)^{d-5} \, .
\end{equation}
The matching contributions that might compete with this can also have a similar form, which we write schematically as
\begin{equation} \label{eq:b2}
    C_{5,\mathrm{Match}} \sim \frac{1}{\Lambda} \left(\frac{M}{\Lambda}\right)^{\delta} \, ,
\end{equation}
where $\delta = 1$ if we can conclude that the neutrino mass must contain a massive parameter in the numerator, and otherwise $\delta = 0$. We distinguish two possible cases where the matching contribution to the Weinberg operator might take this form in our framework:
\begin{enumerate}
    \item \textbf{The non-renormalisable coupling of the $X$ field is a trilinear interaction with a massive coefficient $\mu$.} Recall that our central assumption is that the dominant source of $L$-violation enters through the terms we consider in the tables of App.~\ref{sec:appendixtables}. If this term is such a trilinear interaction, then the neutrino masses must be proportional to $\mu$.
    \item \textbf{In the cases where $X$ is a VLF, the operators providing the $L$-violation feature opposite Dirac-partners of $X$, \textit{i.e.} different chiralities.} In this case, the matching contribution to $C_5$ should be directly proportional to $M$, the $X$ mass, since it is necessary for the chirality flip.
\end{enumerate}

The cases in which the running could contribute sizeably are then those in which the dimension $d$ of Eq.~\eqref{eq:b1} is chosen to match powers of $M$ in Eq.~\eqref{eq:b2}. We consider the two possibilities:
(1) $\delta = 0$ and $d = 5$, and (2) $\delta = 1$ and $d = 6$, which are dominant in the absence of any tuning. These two cases correspond respectively to Eqs.~\eqref{eq:deltad-0} --~\eqref{eq:deltad-1}. In Appendix~\ref{sec:appendixModel} we go on to illustrate these cases in some example UV models.

\section{Ultraviolet examples: dominant contributions} \label{sec:appendixModel}

In Appendix~\ref{sec:matching-and-running-estimates} we discussed the situations in which matching or running dominated the coefficient of either the Weinberg operator or the $d \leq 7$ $B$-violating operators in the SMEFT at low energies. In the following we give concrete UV models that exemplify the three cases highlighted in our work: (1) where the running dominates according to Eqs.~\eqref{eq:deltad-1a} and~\eqref{eq:deltad-1}, (2) where the running dominates according to Eq.~\eqref{eq:deltad-0}, and (3) where the matching dominates. Again, our results here apply equally well to both the neutrino-mass and proton-decay cases, but each of the examples provided are relevant for neutrino masses.

\subsection{Running proportional to \texorpdfstring{$M$ ($\Delta d = 1$)}{M (Delta d=1)}}
\label{sec:deltadeq1}

The UV model we use here to illustrate Weinberg-operator contributions scaling like Eqs.~\eqref{eq:deltad-1a} and~\eqref{eq:deltad-1} was first presented and studied in section 4 of Ref.~\cite{Cai:2014kra}. In our notation, it augments the SM by the scalar leptoquark $\omega_1$ and the vector-like fermion $Q_5$. We point the reader to Ref.~\cite{Cai:2014kra} for a full description of the model; here we are only concerned with the form of the loop integral and the scaling of the operator coefficient with $M$ and $\Lambda$. To make a connection to the framework of our paper, we identify $m_{\omega_1}=\Lambda$ and $m_{Q_5} = M$, \textit{i.e.} we consider the region of parameter space in which $\omega_1$ is heavier than the Dirac fermion $Q_5$. We perform the matching onto the Weinberg operator in the unbroken phase, so there are no complications introduced by the fact that a component of $Q_5$ mixes with the bottom quark. 

\begin{figure}[t]
    \centering
    \includegraphics[width=0.4\linewidth]{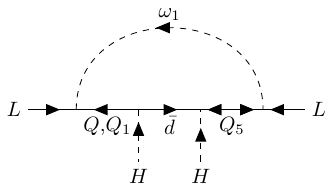}
    \includegraphics[width=0.4\linewidth]{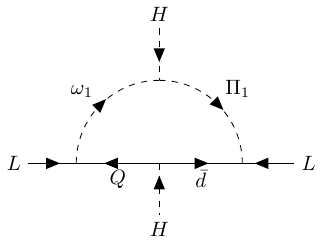}
    \caption{(Left) The neutrino mass diagram for the model presented in Ref.~\cite{Cai:2014kra} for the choice of $Q$ on the first internal fermion line. This is the example model presented in the $\Delta d = 1$ scenario of Sec.~\ref{sec:deltadeq1}. The choice of $Q_1$ is relevant for the $\Delta d > 1$ example model, presented in Sec.~\ref{sec:deltadgtr1}. (Right) The neutrino-mass diagram for the leptoquark model of neutrino masses, as presented in Sec.~\ref{sec:deltadeq0}.}
    \label{fig:vlq-model}
\end{figure}

For simplicity we leave off dimensionless couplings and fix all infrared mass-scales to be equal to $M$. The neutrino-mass diagram is shown in the left panel of Fig.~\ref{fig:vlq-model} for the choice of $Q$ on the first internal fermion line. The structure of the Weinberg-operator coefficient $C_5$ in this case is
\begin{align}
C_5 
&\propto y_b M \int \frac{d^4q}{(2\pi)^4} \frac{1}{q^2}\frac{1}{q^2-\Lambda^2} \frac{1}{q^2-M^2} \\
&\propto \frac{y_b}{16\pi^2} \frac{M}{M^2 - \Lambda^2} \log\left( \frac{M}{\Lambda} \right) \; ,
\end{align}
where $y_b$ is the bottom Yukawa. Taking the limit $M \to \Lambda$ recovers the matching estimate given in Refs.~\cite{deGouvea:2007qla,Gargalionis:2020xvt} for $\mathcal{O}_{3b}$, while the expression we provide in Table~\ref{tab:fermion-neutrino-masses} is recovered when expanding in $M/\Lambda$:
\begin{equation}
C_{5} \propto \frac{y_b}{16\pi^2} \frac{M}{\Lambda^2} \log \left(\frac{M}{\Lambda}\right)+ \mathcal{O}(M^3/\Lambda^3) \; ,
\end{equation}
and this matches the form of Eq.~\eqref{eq:deltad-1}. As argued in App.~\ref{sec:matching-and-running-estimates}, the dominance of the running and the proportionality to $M/\Lambda$ is fixed at the level of the $X$SMEFT by chirality.

\subsection{Mixing between operators of the same dimension (\texorpdfstring{$\Delta d = 0$}{Delta d=0})}
\label{sec:deltadeq0}

As an example UV model providing a running contribution to the Weinberg-operator coefficient like Eq.~\eqref{eq:deltad-0}, we use the so-called \textit{leptoquark model} of neutrino masses~\cite{AristizabalSierra:2007nf,Cai:2014kra,Dorsner:2017wwn}. We note here that the arguments presented below apply equally well to any of the minimal tree-level completions of the dimension-7 $\Delta L=2$ effective operators, as outlined in Ref.~\cite{Cai:2014kra}, including the Zee model~\cite{Zee:1980ai}.

The model extends the SM by two additional leptoquark scalars. In our notation, these are the isosinglet $\omega_1$ and isodoublet $\Pi_1$. The pertinent parts of the Lagrangian of the model are
\begin{equation} \label{eq:b2a}
\mathcal{L}_{\omega_1 \Pi_1} \supset - y_{pr} L^i_p Q^j_r \omega_1^\dagger \epsilon_{ij} - g_{qs} L^i_q \bar{d}_s \Pi_1^j \epsilon_{ij} - \kappa \Pi_1^\dagger H \omega_1 + \mathrm{h.c.} \, ,
\end{equation}
and we present the neutrino-mass diagram in the right panel of Fig.~\ref{fig:vlq-model}. We consider the state of affairs in which $m_{\omega_1}$ is identified with $M$ and $\Pi_1$ is taken to be heavy: $m_{\Pi_1} \sim \Lambda$, with $\kappa \sim \Lambda$ as it is also a UV parameter. As before, the neutrino-mass matrix is proportional to
\begin{align}
C_5 &\propto
iy_b\Lambda \int \frac{d^4 q}{(2\pi)^4} \frac{1}{q^2} \frac{1}{q^2 - \Lambda^2} \frac{1}{q^2 - M^2} \\
  \label{eq:b4}
  &\propto \frac{y_b}{16\pi^2} \frac{\Lambda}{M^2 - \Lambda^2} \log \left( \frac{M}{\Lambda} \right) \, ,
\end{align}
where again $y_b$ is the bottom Yukawa. Taking the limit $M \to \Lambda$ recovers the matching estimate given in Refs.~\cite{deGouvea:2007qla,Gargalionis:2020xvt} for $\mathcal{O}_{3b}$, and expanding in $M/\Lambda$ provides the expression seen in Table~\ref{tab:scalar-neutrino-masses} for $\omega_1$,
\begin{equation}
  C_5 \propto \frac{y_b}{16\pi^2} \frac{1}{\Lambda} \log \left( \frac{M}{\Lambda} \right) + \mathcal{O} (M^2 / \Lambda^2) \, ,
\end{equation}
which matches the form of Eq.~\eqref{eq:deltad-0}.

\subsection{Matching (\texorpdfstring{$\Delta d > 1$}{Delta d>1})}
\label{sec:deltadgtr1}

Here we provide an example model for which matching at the scale $\Lambda$ dominates the expression for the neutrino masses. The arguments presented in App.~\ref{sec:matching-and-running-estimates} suggest that this will happen whenever $\Delta d > 1$. We take as an example a model that extends the SM by the VLF $Q_1$, which we take to have mass $M$, and some additional heavy particle content sufficient to generate the $X$SMEFT operator listed in the corresponding row of Table~\ref{tab:fermion-neutrino-masses}. This additional heavy particle content can be taken to be the scalars $\omega_1$ and $\Pi_1$, that alone are sufficient to generate the one-loop \textit{leptoquark model} contribution discussed above. We will see that the contribution to $C_5$ in the model extended by $Q_1$ is not proportional to a SM Yukawa coupling, and thus can dominate the expression given in Eq.~\eqref{eq:b4}.

The terms relevant for neutrino masses in this mechanism are an extension of those of Eq.~\eqref{eq:b2a}:
\begin{equation}
\mathcal{L}_{\omega_1 \Pi_1 Q_1} \supset (- f_{p} L^i_p Q_{1}^j \omega_1^\dagger + \mathrm{h.c.}) + \mathcal{L}_{\omega_1 \Pi_1} \, .
\end{equation}

The neutrino-mass diagram is shown in the left panel of Fig.~\ref{fig:vlq-model} for the choice of $Q_1$ on the first internal fermion line. We highlight that the fermion $Q_1$ enters with an arrow-preserving propagator, providing a $\sigma \cdot q$ in the numerator of the loop integral:
\begin{align}
I &\propto \Lambda \int \frac{d^4 q}{(2\pi)^4} \frac{\bar{\sigma} \cdot q}{q^2 - M^2} \frac{\sigma \cdot q}{q^2} \left(\frac{1}{q^2 - \Lambda^2}\right)^2 \\
&\propto \frac{\Lambda}{16\pi^2} \left[ \frac{1}{M^2 - \Lambda^2} - \frac{M^2}{(M^2 - \Lambda^2)^2} \log \frac{M^2}{\Lambda^2} \right] \\
&\propto \frac{1}{16\pi^2 \Lambda} + \mathcal{O}(M^2/\Lambda^2) \, ,
\end{align}
whose leading-order piece provides the Weinberg operator coefficient $C_5 \propto 1/(16\pi^2\Lambda)$.

\section{Computation of neutrino mass and nucleon decays} \label{sec:appendixtables}

In this appendix, we provide details of the analysis and results performed for the $\Delta L=2$ and $\Delta B =1$ processes induced by each LSME. In Tabs.~\ref{tab:fermion-neutrino-masses}--\ref{tab:scalar-neutrino-masses-veved}, we consider the contribution of scalar, vectorial and fermionic LSMEs in the (radiative) generation of $m_\nu$, and in Tabs.~\ref{tab:scalar-proton-decay-veved}--\ref{tab:vector-proton-decay} we provide the analogous analysis regarding (radiative) nucleon decays. In the case of scalar LSMEs, we consider both cases where they may or may not develop a VEV, as explained in Sec.~\ref{sec:scalar-veved}.

The structure of the tables of the $\Delta L=2$ processes is the following: in the first two columns, we provide the label of each LSME and its quantum numbers under $G_{\rm SM}$ and Lorentz group. In the third column, we write the combination of operators that will induce Majorana neutrino masses.\footnote{We do not assign $B$ and $L$ to the LSMEs, as only specific combinations of operators lead to the respective violations.} Note that we only write the lowest-dimensional genuine operators from which we obtain the dominant contribution when performing the matching onto the Weinberg operator, as explained in Sec.~\ref{sec:framework}. We explicitly write down the SU(2)$_{\rm L}$ contractions whenever they are uniquely fixed either by genuineness or because they lead to the largest contribution to neutrino masses. The fourth and fifth columns indicate both $\Delta d$ for each LSME and the SMEFT field string onto which these fields match, following the convention of Refs.~\cite{deGouvea:2007qla,Gargalionis:2020xvt}. In the sixth column, we write the (loop-) matching estimate to $[m_\nu]_{pq}$, where the given expressions should be symmetrised by adding the exchange $+(p \leftrightarrow q)$. We also obtain the limits from 1-particle irreducible loop diagrams. Finally, in the last column, we quote the upper bound on the mass $M$ (in TeV), obtained in the limiting case $\Lambda \rightarrow M$. This is done through naive dimensional analysis, assuming a single scale $\Lambda$.

Special attention should be paid to the sixth column, where the suppressions in the generation of $m_\nu$ can be read off. Specifically, one can identify the loop suppressions, $\epsilon =1/(16\pi^2)$, $L=1/(16\pi^2)\log (\frac{\Lambda}{M})$, and $L^\prime=1/(16\pi^2)\log (\frac{v}{M})$,
the insertions of SM Yukawa couplings (which apart from $y_t \simeq 1$, generally suppress the process\footnote{The notation for the SM Yukawa interactions used throughout our tables is as follows: $[Y_u]_{pq} Q^p \bar u^q H+{ \rm H.c.}$, $[Y_d]_{pq} Q^p \bar d^q H^\dagger+{\rm H.c.}$, and $[Y_e]_{pq} L^p \bar e^q H^\dagger +{\rm H.c.}$. Our results are presented on the up-quark mass basis, where the CKM matrix appears in the down-quark sector, $Y_d = V^\dagger \cdot Y_d^D$. In this basis, we identify $[Y_u]_{pq} = [y_u]_p \,\delta_{pq}$, $[Y_e] = [y_e]_p \delta_{pq}$, and $[Y_d]_{pq} = [y_d]_q \, V_{qp}^*$, with $[y_{u,d,e}] \in \mathbb{R}$ denoting the diagonal entries of the SM Yukawa matrices, and there is no sum over the indices $p,q$. We note that we work with the neutrinos as flavour eigenstates.\label{foot:basis}}), the $W$-boson exchanges (where we take $g \simeq 0.6$), and the structure of the diagram (loop) as determined by the explicit contractions between the beyond-the-SM (BSM) couplings $y, c$, and the SM ones.

On the other hand, the structure of the tables of $\Delta B=1$ processes is the following: in the first two columns, we provide the label of each LSME and its quantum numbers under $G_{\rm SM}$ and Lorentz group. In the third column, we write the combination of operators that will induce $\Delta B =1$ nucleon decay. As for the $\Delta L=2$ tables, we only write the lowest-dimensional genuine operators from which we obtain the dominant contribution when performing the matching onto the BNV SMEFT operators. The fourth and fifth columns indicate the $\Delta d$ of the process and the resulting $\Delta B =1$ SMEFT field string using the notation of Ref.~\cite{Gargalionis:2024nij}, respectively. In the sixth column, unlike the $\Delta L=2$ case where the (loop-)matching is performed uniquely to the dimension-5 Weinberg operator, for nucleon decay we have four (six) dimension-6 (7) BNV SMEFT operators, leading to a more complex algorithm to find the leading contribution, where we also provide the flavour indices of such operators as they play a significant role in determining the dominant nucleon decay channel~\cite{Gargalionis:2024nij}. Finally, in the seventh column, again, unlike for neutrino masses, we explicitly indicate the nucleon decay channel induced by the interplay of the operators displayed in the third column from which we derive the quoted bound on $M$ in TeV provided in the last column in the limiting case $\Lambda \rightarrow M$.\footnote{It is important to note that the displayed nucleon decay channel does not imply exclusivity; rather, it is the channel used to extract the bound based on projected sensitivities from Hyper-K. Crucially, if nucleon decay is observed, irrespective of the detected channel, it could originate from any of the LSMEs under study.}

In Tabs.~\ref{tab:scalar-neutrino-masses-veved} and \ref{tab:scalar-proton-decay-veved} we show the possibility in which the neutral components of the LSME scalars acquire a VEV. For nucleon decay (Tab.~\ref{tab:scalar-proton-decay-veved}), we perform the tree-level matching directly onto the dimension-6 $\Delta B = 1$ LEFT WC $L^{S,XY}_{q_1q_2q_3}$, displayed in the fourth column, following the notation of Refs.~\cite{Beneito:2023xbk, Jenkins:2017jig}. We explicitly denote flavour symmetries using curly (square) brackets to indicate pairs of flavour indices that are symmetric (antisymmetric) under permutation whenever multiple copies of a field appear in an operator. A word of caution is in order for non-renormalisable operators that involve either more than two fermionic fields of the same type or two pairs of identical fermions, as the resulting flavour structure can be highly non-trivial, as mentioned in Sec.~\ref{sec:flavoursymmetries}.

Let us also mention some $\mathcal{O}(1)$ factors that we have neglected in our analysis. First, we do not take into account the colour factors arising from LSMEs charged under SU(3)$_{\rm C}$ circulating within a loop, nor do we consider the effect of renormalisation group
running from the scale $M$ down to the nuclear scale where the $\Delta B=1$ observables are computed~\cite{Beneito:2023xbk}.\footnote{We also neglect the RG effects for the Weinberg operator~\cite{Casas:1999tg,Antusch:2001ck}.} Furthermore, we also neglect the factors arising from both Schouten and Fierz identities, even for those LSMEs that generate nucleon decay and Majorana neutrino masses at tree level. We refer the reader to Refs.~\cite{deBlas:2017xtg,Li:2022abx} for a detailed analysis of the particles that fall into the category $T_{\Delta (B-L)=0,\,2}$. For the generation of Majorana neutrino masses, we assume an approximately diagonal CKM matrix $V \approx \mathbb{1}$ and therefore do not display it in Tabs.~\ref{tab:fermion-neutrino-masses}--\ref{tab:scalar-neutrino-masses-veved}. However, a CKM matrix element (or its complex conjugate) is implicitly present whenever a $[y_d]$ contribution appears in the sixth column, see Footnote \ref{foot:basis}. This contrasts with $\Delta B=1 $ processes, where CKM elements are explicitly included in Tabs.~\ref{tab:scalar-proton-decay-veved}--\ref{tab:vector-proton-decay}, given their crucial role, as discussed in Sec.~\ref{sec:flavoursymmetries}. The reason is simply that, for Majorana neutrino masses, pairs of quark fields are usually annihilated via SM Yukawa couplings as they are not present in the dimension-5 Weinberg operator, whereas for nucleon decay they must appear explicitly in the final SMEFT operator to which our operator is matched. Since nucleon decay is mediated by light-family quarks, there is a fundamental trade-off between inserting large Yukawa couplings and the correct external light-family quark flavours, which is parametrised by the CKM matrix $V$. Alternatively, one could set all the SM Yukawa couplings in the loop matching expressions to the first family and take the diagonal elements of $V$. However, typically this does not yield the leading contribution to nucleon decay.

Finally, let us briefly comment on the presence of SMEFT field strings involving a covariant derivative in the case of the vector-like fermions $E$, $D$, $Q_1$, and $\Delta_1$ for $\Delta L=2$ operators, and the scalar $\mathcal{S}_2$ and the vector $\mathcal{U}_5$ for $\Delta B=1$ operators. Such covariant derivatives arise from the closure of identical SM fermions, which carries a factor of momentum from the propagator $\psi^\dagger \slashed{D}\psi $. For a detailed analysis of such operators closure, see Sec.~3.2 of Ref.~\cite{Gargalionis:2020xvt}. For the analysis of nucleon decay, except for $\Omega_4$ and $\varphi$, all LSMEs are matched onto a dimension-$d$ BNV SMEFT field string with $d \leq 9$. Consequently, the algorithm presented in Ref.~\cite{Gargalionis:2024nij} can be directly applied to derive a bound, incorporating updated sensitivities from Hyper-K; for definiteness, we use $\tau_p = 10^{35}$ years. The fields mentioned above necessarily involve SMEFT field strings of dimension-10, which then we match onto dimension-6 SMEFT operators. For those we use an updated version of the algorithm in Ref.~\cite{Gargalionis:2024nij}. Out of the many field strings arising at this mass dimension, we select the field strings $\bar e^\dagger QQ Q^\dagger \bar u^\dagger \bar u^\dagger H^\dagger$, and $ L Q Q \bar d^\dagger \bar d^\dagger \bar d  H$, which we refer as $\mathcal{O}_{\Omega_4}$ and $\mathcal{O}_\varphi$ in Tab.~\ref{tab:scalar-proton-decay}, as they provide the leading contribution. For completeness, the UV model that generates these operators at the tree level requires three additional exotic BSM particles beyond $\varphi$ or $\Omega$, which is in conflict with the goal of focusing on simple UV completions with minimal BSM content. In the case of $\varphi$, the lowest-dimensional genuine $\varphi$SMEFT operators leading dominantly to nucleon decay and neutrino masses contain an explicit SU(2)$_{\rm L}$ antisymmetric contraction between both Higgs doublets, namely $\epsilon_{ij}\varphi^i H^j \sim \mathcal{S}_1$. Finally, let us mention the special case of $\mathcal{B}_1$ in $\Delta L=2$ processes, where we refer the reader to Ref.~\cite{Gargalionis:2020xvt}. For such LSME, an additional $W$ boson exchange is needed to avoid a non-trivial cancellation, leading to a further suppressed loop matching contribution.
    

\begin{sidewaystable}[p]
\centering
{
\setlength{\arrayrulewidth}{0.2mm}
\setlength{\tabcolsep}{7pt}
\renewcommand{\arraystretch}{1.9}
\resizebox{\textwidth}{!}{
\begin{tabular}{ccccccc}
         \hline \hline
         LSME & $G_{\rm SM}$ & $\mathcal{L}_{\Delta L = 2}$ & $\Delta d$ & Op. & $[m_{\nu}]_{pq}$ & Upper Bound (TeV)\\ \hline
         
         $ N $ & $ (1,1,0)_F$ & $ y_{p} (NL_{p}) H$ + $M (NN)$ & $-2$ & $\mathcal{O}_1$ & $y_{p}y_{q}\frac{v^2}{M}$ & $10^{12}$
         \\

         $ \Sigma $ & $ (1,3,0)_F$ & $y_{p} (\Sigma L_{p}) H$ + $M(\Sigma \Sigma)$ & $-2$ & $\mathcal{O}_1$ &  $y_{p}y_{q}\frac{v^2}{M}$ & $10^{12}$
         \\

         $\Sigma_1 $ & $ (1,3,-1)_F$ & $ y_{p}  (\bar{\Sigma}_{1} L_p) H^\dagger$ + $c_{q} (\Sigma_1 L_q)HHH$ & $1$ & $\mathcal{O}_{1}^{\prime}$  &  $y_{p} c_{q} \frac{M}{\Lambda} \frac{v^2}{\Lambda} \; L$ & 
         $5 \cdot 10^9$
         \\

        $ Q_7 $ & $ (3,2,7/6)_F$ & $ y_{s} (Q_7^{\dagger}   \Bar{u}_s^\dagger) H$ + $c_{pqr} (L_{pi} L_{qj}) (\Bar{Q}_{7}^{\dagger i} Q_{r}^{\dagger j})$ & $1$ & $\mathcal{O}_{4a}$ & $y_rc_{pqr} [y_u]_{r} \frac{M}{\Lambda} \frac{v^2}{\Lambda} \; L$ &
        $5 \cdot 10^9$ \\

        $E $ & $ (1,1,-1)_F$ & $y_{s} (\Bar{E}^\dagger L^\dagger_s)H$ + $c_{pqr} (L_{p}^{i} L_{q}^{j})(L_{r}^{k} \bar{E}) H^{l} \epsilon_{ik} \epsilon_{jl}$ & $2$ & $\mathcal{O}_{D5b}$ & $ y_{r} c_{pqr} \frac{v^2}{\Lambda} \; \epsilon$ & 
        $5 \cdot 10^9$
        \\

        $ D $ & $ (3,1,-1/3)_F$ & $y_{s} (\Bar{D}^\dagger Q_{s}^\dagger) H$ + $c_{pqr}(L_p^i \bar{D})(L_q^j Q_r^k) H^l \epsilon_{ik} \epsilon_{jl}$ & 2 & $\mathcal{O}_{D8a}$ & $y_{r}c_{pqr} \frac{v^2}{\Lambda} \; \epsilon$ & 
        $5 \cdot 10^9$
        \\

        $T_1 $ & $ (3,3,-1/3)_F $& $y_{s} (\Bar{T}_1 Q_s) H^\dagger$ + $c_{\{pq\}r} (L^\dagger_p L^\dagger_q)(T_1 \bar{u}_{r})$ & $1$ & $\mathcal{O}_{4a}$ & $y_rc_{\{pq\}r}  [y_u]_{r} \frac{M}{\Lambda} \frac{v^2}{\Lambda} \; L$ & 
        $5 \cdot 10^9$  \\

        $ Q_1 $ & $ (3,2,1/6)_F$ & $y_{s}( Q_1^\dagger \Bar{d}_s^\dagger) H$ + $c_{pqr} (L_{p}^{i}Q_1^{j})(L_{q}^{k} \Bar{d}_{r}) H^{l} \epsilon_{ij} \epsilon_{kl}$ & $2$ &  $\mathcal{O}_{D9a} $ & $y_r c_{pqr} \frac{v^2}{\Lambda} \; \epsilon$ & 
        $5 \cdot 10^9$
        \\

        $\Delta_1 $ & $ (1,2,-1/2)_F$ & $y_{s} (\Delta_1^\dagger  \Bar{e}_s^\dagger) H$ + $c_{pqr} (L_{p}^{i}\Delta_1^{j}) (L_{q}^{k} \bar{e}_{r}) H^{l} \epsilon_{ij} \epsilon_{kl}$& $2$ &  $\mathcal{O}_{D6a}$ & $y_r c_{pqr}  \frac{v^2}{\Lambda} \; \epsilon$ & 
        $5 \cdot 10^9$
        \\
        
        $ Q_5 $ & $ (3,2,-5/6)_F$ & $y_{s}(Q_5 \Bar{d}_s) H$ + $c_{pqr} (L^{i}_{p}\Bar{Q}^{j}_5) (L^{k}_{q} Q^{l}_{r})\epsilon_{ij} \epsilon_{kl}$ & $1$ & $\mathcal{O}_{3b}$ & $y_rc_{pqr} [y_d]_{r} \frac{M}{\Lambda} \frac{v^2}{\Lambda} \; L$ &
        $10^8$ \\

        $T_2 $ & $ (3,3,2/3)_F$& $y_{s} (\Bar{T}_2 Q_s) H$ + $c_{pqr} (T_2 L_{p}) (L_{q} \Bar{d}_{r})$ & $1$ & $ \mathcal{O}_{3b}$ & $y_rc_{pqr} [y_d]_{r} \frac{M}{\Lambda} \frac{v^2}{\Lambda} \; L$ & $10^8$ \\

        $\Delta_3 $ & $ (1,2,-3/2)_F$ & $y_{s} (\Delta_3 \Bar{e}_s) H $ + $c_{pqr} (L_{p}^{i} L_{q}^{j})(L^{k}_{r}\bar{\Delta}_3^{l}) \epsilon_{ik} \epsilon_{jl}$ & $1$ & $\mathcal{O}_2$ & $y_r c_{pqr} [y_{e}]_{r} \frac{M}{\Lambda} \frac{v^2}{\Lambda} \; L$ & $5 \cdot 10^7$ \\

        $ U $ & $ (3,1,2/3)_F$ & $y_{s} (\bar{U} Q_s) H$ + $c_{pqr} (U L_{p}) (L_{q} \Bar{d}_{r})$ & $1$ & $\mathcal{O}_{3a}$ &  $y_r c_{pqr} [y_d]_{r} g^2 \frac{M}{\Lambda} \frac{v^2}{\Lambda} \; L^2$ & $5 \cdot 10^5$ \\

        \hline \hline
\end{tabular}
}
\caption{\label{tab:fermion-neutrino-masses} Majorana neutrino mass generation for vector-like fermions. Unless explicitly symmetric in $p,q$, the expressions given should be extended by $+ (p \leftrightarrow q)$ terms. The non-renormalisable operators of dimension $d>4$, listed in the third column, include a suppression factor $1/\Lambda^{d-4}$. The new couplings $y$ and $c$ are assumed to be dimensionless. The upper bound in the last column is derived assuming $M \leq \Lambda$, where $y$ and $c$ have been set to unity. While colour contractions are unique and therefore not shown explicitly, we follow Ref.~\cite{Dreiner:2008tw} to denote Lorentz contractions, which are explicitly indicated in parentheses $(\psi \psi) \equiv \psi^\alpha \psi^\beta \epsilon_{\alpha \beta}$, and $(\psi^\dagger \psi^\dagger) \equiv \psi^{\dagger}_{\dot \alpha} \psi^{\dagger}_{\dot \beta} \epsilon^{\dot \alpha \dot \beta}$. Additionally, SU(2)$_{\rm L}$ indices are displayed according to the criteria explained in App.~\ref{sec:appendixtables}. Flavour indices $p,q$ are fixed in the matching onto the LEFT Wilson coefficients $[m_\nu]_{pq}$, whereas flavour indices $r,s$ are assumed to be summed over. In our notation, the Weyl spinor $X$ combines with $\bar X^\dagger$ to form a Dirac spinor. The fields $N$ and $\Sigma$ are assumed to be Majorana, requiring no Dirac partner. The term ``+ H.c.'' is understood to be included in the third column. Loop suppression factors in the sixth column are defined as $\epsilon =1/(16\pi^2)$, and $L=1/(16\pi^2)\log (\frac{\Lambda}{M})$.}
}
\end{sidewaystable}

\begin{sidewaystable}[p]
\centering
{
\setlength{\arrayrulewidth}{0.2mm}
\setlength{\tabcolsep}{7pt}
\renewcommand{\arraystretch}{1.9}
\resizebox{\textwidth}{!}{
\begin{tabular}{ccccccc}
         \hline \hline
         LSME & $G_{\rm SM}$ & $\mathcal{L}_{\Delta L = 2}$ & $\Delta d$ & Op. & $[m_{\nu}]_{pq}$ & Upper Bound (TeV)\\ \hline
        
        $\Xi_1 $ & $ (1,3,1)_S$ & $y_{\{pq\}} \: \Xi_1 (L_p L_q)$ + $\mu  {\Xi_1^\dagger} H H$ & $-2$ & $\mathcal{O}_1$ &  $y_{\{pq\}} \frac{\mu}{M} \frac{v^{2}}{M}$ & $10^{12}$
        \\
        
        $\Xi $ & $ (1,3,0)_S$ & $ \mu \, \Xi H^{\dagger} H$ + $c_{\{pq\}} \Xi^{\{ij\}} (L^k_p L^l_q) H^m H^n \epsilon_{km} \epsilon_{il} \epsilon_{jn}$ & $1$ & $\mathcal{O}_1^\prime$ & $c_{\{pq\}} \left( L + \frac{v^2}{\Lambda^2}\right) \frac{\mu}{\Lambda} \frac{v^{2}}{\Lambda}$ & $5 \cdot 10^9$
        \\

        $\Theta_3 $ & $ (1,4,3/2)_S$ & $y \Theta_3^\dagger H H H$ + $c_{\{pq\}} \Theta_3 (L_p L_q) H^\dagger$ & $0$ & $\mathcal{O}_1^\prime$ & $ y \,c_{\{pq\}} \left( L + \frac{v^2}{\Lambda^2}\right) \frac{v^2}{\Lambda} $ & $5 \cdot 10^9$
        \\

        $\Pi_7 $ & $ (3,2,7/6)_S$ & $y_{pr} \Pi_7 (\Bar{u}_{r} L_{p})$ + {$c_{qs} \Pi^{\dagger}_{7i} L^{j}_{q} \bar{u}^{\dagger}_{s} (DH)^{i} H^{l} \epsilon_{jl}$} & $2$ & $\mathcal{O}_{D12a}$  &  $y_{pr}c_{qr} \frac{v^2}{\Lambda} \; \epsilon$ & 
        $5 \cdot 10^9$
        \\
        
        $\omega_1 $ & $ (3,1,-1/3)_S$& $y_{pr} \omega_1^\dagger (L_{p} Q_{r})$ + $c_{qs} \omega_1 (L_{q} \bar{d}_{s}) H $ & $0$ & $\mathcal{O}_{3b}$ &  $ y_{pr}c_{qr} [y_d]_{r} \frac{v^2}{\Lambda} \;L$ & $10^8$
        \\
        
        $\zeta $ & $ (3,3,-1/3)_S$ & $y_{pr} \zeta^\dagger (L_{p} Q_{r})$ + $c_{qs} \zeta (L_{q} \bar{d}_{s}) H $ & $0$ & $\mathcal{O}_{3b}$ & $ y_{pr}c_{qr} [y_d]_{r} \frac{v^2}{\Lambda}\;L$ & $10^8$
        \\

        $\Pi_1 $ & $ (3,2,1/6)_S$ & $ y_{pr} \Pi_1 (L_{p} \bar{d}_{r})$ + $c_{qs} {\Pi_1^\dagger}_{i} (L_{q}^{j} Q_{s}^{i}) H^{l} \epsilon_{jl}$ & $0$ & $ \mathcal{O}_{3b}$ & $ y_{pr}c_{qr}[y_d]_{r} \frac{v^2}{\Lambda}\;L$ & $10^8$
        \\

        $\varphi $ & $ (1,2,1/2)_S$ & $y_{rp} \varphi^\dagger (\bar{e}_r L_p)$ + $c_{[qs]} \varphi^i H^j (L^k_q L_s^l) \epsilon_{ij} \epsilon_{kl}$ & $0$ & $\mathcal{O}_{2}$ & $  y_{sq}c_{[sp]} [y_e]_s\frac{v^{2}}{\Lambda} \; L$ & 
        $ 5 \cdot 10^{7}$
        \\

        $\Theta_1 $ & $ (1,4,1/2)_S$ & $y \Theta_1^{\dagger} H H^\dagger H$ + $c_{pq} \Theta_1^{\{ijk\}} (L_p^l L_q^m) H^n H^\dagger_l H^o \epsilon_{im} \epsilon_{jn} \epsilon_{ko}$ & $2$ & $\mathcal{O}_1^{\prime\prime}$ & $y \, c_{pq} \left[ \epsilon^2 + \left(\frac{v^2}{\Lambda^2}\right)^2 \right] \frac{v^2}{\Lambda}$ & 
        $5 \cdot 10^7$
        \\

        $\mathcal{S}_1 $ & $ (1,1,1)_S$ & $y_{[pr]}\mathcal{S}_1 (L_p L_r)$ + $c_{qs} \mathcal{S}_1^{\dagger} (L_{q} \bar{e}_{s}) H$ & $0$ & $\mathcal{O}_2$ & $y_{[pr]} c_{qr}[y_e]_{r} \frac{v^2}{\Lambda} \;L$ & $ 5 \cdot 10^{7}$
        \\

        $\Omega_4 $ & $ (6,1,4/3)_S$ & $y_{\{rs\}} {\Omega_4^{\dagger}} (\bar{u}^{\dagger}_{r} \bar{u}^{\dagger}_{s})$ + $c_{pqtu} \Omega_4 (L_{p} L_{q})^{\{ij\}} (Q^{\dagger}_{t} Q^{\dagger}_{u})_{ij}$ & $2$ & $\mathcal{O}_{12a}$ & $y_{\{rs\}}c_{pqrs}  [y_u]_{r} [y_{u}]_{s} \frac{v^2}{\Lambda} \; \epsilon^2$ & $5 \cdot 10^7$
        \\

        $\Upsilon $ & $ (6,3,1/3)_S$ & $y_{\{rs\}} \Upsilon
        (Q^{\dagger}_r Q^{\dagger}_s)$ + $ c_{\{pq\}\{tu\}} \Upsilon^{\dagger}(L_{p}L_{q}) (\Bar{u}^{\dagger}_{t}\Bar{u}^{\dagger}_{u})$ & 2 & $\mathcal{O}_{12a}$ & $y_{sr}c_{\{pq\}\{rs\}} [y_u]_{r} [y_{u}]_{s} \frac{v^2}{\Lambda} \; \epsilon^2$ & $5 \cdot 10^7$
        \\

        $\Phi $ & $ (8,2,1/2)_S$ & $y_{rs} \Phi^\dagger (Q^\dagger_r \Bar{u}_{s}^\dagger)$ + $c_{pqtu} \Phi^i (L_{p}^{j} L_{q}^{k}) (Q^{\dagger}_{tk} \Bar{u}^{\dagger}_{u}) \epsilon_{ij}$ & $2$ & $\mathcal{O}_{12a}$ & $y_{sr}c_{pqrs} [y_u]_{r} [y_{u}]_{s} \frac{v^2}{\Lambda} \; \epsilon^2$ & 
        $5 \cdot 10^7$
        \\

        $\Omega_2 $ & $ (6,1,-2/3)_S$ & $y_{\{rs\}} \Omega_2 (\Bar{d}_{r}\Bar{d}_{s})$ + $c_{pqtu} \Omega^{\dagger}_{2} (L_{p}^{i}Q_{t}^{k}) (L_{q}^{j} Q_{u}^{l}) \epsilon_{ik} \epsilon_{jl}$ & $2$ & $\mathcal{O}_{11b}$ & $y_{\{rs\}}c_{pqrs} [y_d]_{r} [y_{d}]_{s} \frac{v^2}{\Lambda} \; \epsilon^2$ & $10^4$
        \\
        
        $\omega_4 $ & $ (3,1,4/3)_S$ & $y_{rs} \omega_4 (\Bar{e}_{r}\Bar{d}_{s})$ + $c_{pqrs} \omega^{\dagger}_4 (L_{p}^{i} L_{q}^{j}) (L_{r}^{k} Q_{s}^{l})\epsilon_{ik} \epsilon_{jl}$ & $2$ & $\mathcal{O}_{10}$ &  $y_{rs}c_{\{pqr\}s}  [y_{e}]_{r} [y_{d}]_{s} \frac{v^2}{\Lambda} \; \epsilon^2$ & $10^4$
        \\

        $\mathcal{S}_2 $ & $ (1,1,2)_S$ & $ y_{\{rs\}}  \mathcal{S}_2^{\dagger} (\Bar{e}_r \Bar{e}_s)$ + $c_{[pq][tu]} \mathcal{S}_2 (L^{i}_p L^{j}_t) (L^{k}_q L^{l}_u) \epsilon_{ij}\epsilon_{kl}$ & $2$ & $\mathcal{O}_9$ &  $y_{\{rs\}}c_{pqrs}  [y_{e}]_{r} [y_{e}]_{s} \frac{v^2}{\Lambda} \; \epsilon^2$ & $5 \cdot 10^5$
        \\

        $\Omega_1 $ & $ (6,1,1/3)_S$ & $y_{rs} \Omega_1^{\dagger} (\bar{u}^{\dagger}_{r} \bar{d}^{\dagger}_{s})$ + $c_{pqtu} \Omega_1 (L_{p} \bar{d}_{t}) (L_{q} \bar{d}_{u})$ & $2$ & $\mathcal{O}_{17}$ & $y_{ru}c_{pqtu}  [y_{d}]_r [y_{u}]_r g^{2} \frac{v^2}{\Lambda} \; \epsilon^3$ & $10^3$
        \\

        $\omega_2 $ & $ (3,1,2/3)_S$ & $ y_{[rs]} \omega^{\dagger}_{2} (\Bar{d}_{r}\Bar{d}_{s})$ + $c_{pqtu} \omega_{2} (L^{i}_{p} Q^{j}_{t}) (L^{k}_{q} Q^{l}_{u}) \epsilon_{ij} \epsilon_{kl}$ & $2$ &  $\mathcal{O}_{11b} $ & $y_{[rs]}c_{pqrs}  [y_d]_{r} [y_{d}]_{s} \frac{v^2}{\Lambda}\; \epsilon^2$ & $5 \cdot 10^2$
        \\
        
        \hline \hline
\end{tabular}
}
\caption{\label{tab:scalar-neutrino-masses}
Same as Tab.~\ref{tab:fermion-neutrino-masses} for Majorana neutrino masses generated by scalar LSMEs.}
}
\end{sidewaystable}

\begin{sidewaystable}[p]
\centering
{
\footnotesize
\setlength{\arrayrulewidth}{0.2mm}
\setlength{\tabcolsep}{7pt}
\renewcommand{\arraystretch}{1.9}
\resizebox{\textwidth}{!}{
\begin{tabular}{ccccccc}
    \hline \hline

    LSME & $G_{\rm SM}$ & $\mathcal{L}_{\Delta L = 2}$ & $\Delta d$ & Op. & $[m_{\nu}]_{pq}$ & Upper Bound (TeV)\\ \hline
    
    $\mathcal{U}_2 $ & $ (3,1,2/3)_V$ & $y_{pq} \mathcal{U}_2 Q^\dagger_p L_q $ + $c_{rs} \mathcal{U}_2^\dagger L_r \bar u_s^\dagger H$ & 0 & $\mathcal{O}_{4a}$ & $y_{rq} c_{pr} [y_u]_r \frac{v^2}{\Lambda} \; L$ & $ 5 \cdot 10^9$ \\

    $\mathcal{X} $ & $ (3,3,2/3)_V$ & $y_{pq} \mathcal{X} Q^\dagger_p L_q$ + $c_{rs} \mathcal{X}^\dagger L_r \bar u_s^\dagger H$ & 0 & $\mathcal{O}_{4a}$ & $y_{rq} c_{pr} [y_u]_r \frac{v^2}{\Lambda} \; L$ & $5 \cdot 10^9$ \\

    $\mathcal{Q}_1 $ & $ (3,2,1/6)_V$ & $y \mathcal{Q}_1^\dagger
    \bar u_p^\dagger L_q$ + $c_{rs} \mathcal{Q}_1 L_r Q^\dagger_s H$ & 0 & $\mathcal{O}_{4a}$ & $y_{rq} c_{pr} [y_u]_r \frac{v^2}{\Lambda} \; L$ & $5 \cdot 10^9$\\

    $\mathcal{L}_1 $ & $ (1,2,1/2)_V$ & $\mu \mathcal{L}_1^\dagger DH$ +  $c_{pq} \mathcal{L}_1 L_p \bar e^\dagger_q HH$ & 1 & $\mathcal{O}_{D3}$ &  $ c_{pq}[y_e]_q \frac{\mu}{\Lambda} \frac{v^2}{\Lambda} \; L$ & $ 5 \cdot 10^9$ \\

    $\mathcal{W}_1 $ & $ (1,3,1)_V$ & $y\mathcal{W}_1^\dagger H DH$ + $c_{pq} \mathcal{W}_1 L_p \bar e^\dagger_q H$ & 0 & $\mathcal{O}_{D3}$ & $ y c_{pq} [y_e]_q \frac{v^2}{\Lambda} \; L$ & $5 \cdot 10^9$ \\
    
    $\mathcal{L}_3 $ & $ (1,2,-3/2)_V$ & $y_{pq} \mathcal{L}_3^\dagger \bar e_p^\dagger L_q$ + $c \mathcal{L}_3 D H H H $ & 0 & $\mathcal{O}_{D3}$ & $ y_{pq} c [y_e]_q \frac{v^2}{\Lambda} \; L$ & $ 5 \cdot 10^9$ \\

    $ \mathcal{Y}_1 $ & $ (\bar 6,2,1/6)_V$ & $y_{pq} \mathcal{Y}_1^\dagger Q_p^\dagger \bar d_q$ + $c_{rstu} \mathcal{Y}_1^i \bar u^\dagger_r Q_s^j L_t^k L_u^l \epsilon_{ik} \epsilon_{jl}$ & 2 & $\mathcal{O}_{14b}$& $ y_{rs} c_{rspq} [y_u]_r [y_d]_s \frac{v^2}{\Lambda} \; \epsilon^2$ & $10^6$\\

    $ \mathcal{Y}_5 $ & $ (\bar 6,2,-5/6)_V$ & $ y_{pq} \mathcal{Y}_5  Q_p \bar u^\dagger_q$ + $ c_{rstu} \mathcal{Y}_{5i}^{\dagger} L^i_r L^j_s Q_{tj}^{\dagger} \bar d_u $ & 2 & $\mathcal{O}_{14b}$ & $ y_{sr} c_{pqrs} [y_u]_r [y_d]_s \frac{v^2}{\Lambda}\; \epsilon^2$ & $10^6$\\

    $\mathcal{G}_1 $ & $ (8,1,1)_V $ & $y_{pq} \mathcal{G}_1^\dagger \bar u^\dagger_p \bar d_q$ + $c_{rstu} \mathcal{G}_1  L_r^i L_s^j Q^\dagger_{it} Q_u^k \epsilon_{jk}$ & 2 & $\mathcal{O}_{14b}$ & $ y_{rs} c_{pqrs} [y_u]_r [y_d]_s \frac{v^2}{\Lambda} \; \epsilon^2$ & $10^6$\\

    $\mathcal{H} $ & $ (8,3,0)_V$ & $y_{pq} \mathcal{H} Q^\dagger_p Q_q$ + $c_{rstu} \mathcal{H} L_r L_s \bar u^\dagger_t \bar d_u$ & 2 & $\mathcal{O}_{14b}$ & $ y_{rs} c_{pqrs} [y_u]_r [y_d]_s \frac{v^2}{\Lambda} \; \epsilon^2$ & $10^6$\\

    $\mathcal{W} $ & $ (1,3,0)_V $ & $y_{pq} \mathcal{W}Q^\dagger_p Q_q$ + $c_{rstu} \mathcal{W} L_r L_s \bar u^\dagger_t \bar d_u$ & 2 & $\mathcal{O}_{14b}$ & $ y_{rs} c_{pqrs} [y_u]_r [y_d]_s \frac{v^2}{\Lambda} \; \epsilon^2$ & $10^6$\\

    $ \mathcal{Q}_5 $ & $ (3,2,-5/6)_V$ & $y_{pq} \mathcal{Q}_5 \bar e_p Q^\dagger_q$ + $ c_{rstu}\mathcal{Q}_5^\dagger \bar u_r^\dagger L_s L_t L_u$ & 2 & $\mathcal{O}_{13}$& $y_{sr}c_{rspq} [y_u]_r [y_e]_s \frac{v^2}{\Lambda} \; \epsilon^2$ & $5 \cdot 10^5$\\

    $\mathcal{U}_5 $ & $ (3,1,5/3)_V$ & $y_{pq} \mathcal{U}_5^\dagger \bar e_p \bar u_q^\dagger$ + $c_{rstu} \mathcal{U}_5 Q^\dagger_r L_s L_t L_u$ & 2 & $\mathcal{O}_{13}$ & $y_{sr}c_{rspq} [y_u]_r [y_e]_s \frac{v^2}{\Lambda} \; \epsilon^2$ & $5 \cdot 10^5$\\

    $\mathcal{G} $ & $ (8,1,0)_V$ & $y_{pq} \mathcal{G} Q^\dagger_p Q_q$ + $c_{rstu} \mathcal{G} L_r L_s \bar u^\dagger_t \bar d_u$ & 2 & $\mathcal{O}_{14a}$ & $ y_{rs} c_{pqrs} [y_u]_r [y_d]_s  g^2\frac{v^2}{\Lambda} \; \epsilon^3$ & $10^3$\\

    $\mathcal{B} $ & $ (1,1,0)_V$ & $y_{pq} \mathcal{B} Q^\dagger_p Q_q$ + $c_{rstu} \mathcal{B} L_r L_s \bar u^\dagger_t \bar d_u$ & 2 & $\mathcal{O}_{14a}$ & $ y_{rs} c_{pqrs} [y_u]_r [y_d]_s  g^2\frac{v^2}{\Lambda} \; \epsilon^3$ & $10^3$\\
    
    $\mathcal{B}_1 $ & $ (1,1,1)_V$ & $y_{pq}\mathcal{B}_1^\dagger \bar u_p^\dagger \bar d_q$ +  $c_{rs} \mathcal{B}_1 L_r \bar e^\dagger_s H$ & 0 & $\mathcal{O}_8$  & $ y_{ss}c_{pq} [y_e]_q [y_u]_s [y_d]_s g^2\frac{v^2}{\Lambda} \; L^3$ & $ 50$ \\
    \hline \hline
\end{tabular}
}
\caption{\label{tab:vector-neutrino-masses}
Same as Tab.~\ref{tab:fermion-neutrino-masses} for Majorana neutrino masses generated by vector LSMEs. Explicit Lorentz contractions involving sigma matrices $\sigma$ or $\bar \sigma$ are not shown, as they are uniquely determined for renormalisable operators but remain ambiguous for non-renormalisable ones.
}
}

\end{sidewaystable}


\begin{sidewaystable}[p]
\centering
{
\setlength{\arrayrulewidth}{0.2mm}
\setlength{\tabcolsep}{7pt}
\renewcommand{\arraystretch}{1.9}

\resizebox{\textwidth}{!}{
\begin{tabular}{cccccc}
        \hline \hline
        LSME & $G_{\rm SM}$ & $\mathcal{L}_{\Delta L = 2}$ & $[m_{\nu}]_{pq}$ & Upper Bound (TeV) & $\langle X^0 \rangle$ (TeV) \\ \hline
        
        $ \Xi_1 $ & $(1,3,1)_S$ & $\mu  \Xi_1^\dagger H H$ + $c_{\{pq\}} \: \Xi_1 (L_p L_q)$ &
        $ c_{\{pq\}} \left( \mu \; \frac{v^2}{M^2} \right)$ & $ 5 \cdot 10^{12}$ & $5 \cdot 10^{-15}$
        \\

        $ \Xi$ & $(1,3,0)_S$ & $ \mu \Xi H^{\dagger} H$ + $c_{\{pq\}} \Xi^{\{ij\}} (L^k_p L^l_q) H^m H^n \epsilon_{km} \epsilon_{il} \epsilon_{jn}$ & $c_{\{pq\}} \; \frac{v^2}{\Lambda^2} \left( \mu \;\frac{v^2}{M^2} \right)$ &
        $5 \cdot 10^3$ & $6 \cdot 10^{-6}$
        \\

        $ \Theta_3 $ & $(1,4,3/2)_S$ & $y {\Theta_3^\dagger} H H H$ + $c_{\{pq\}} \Theta_3 (L_p L_q) H^{\dagger}$ & $c_{\{ pq \}} \; \frac{v}{\Lambda} \left( y \; \frac{v^3}{M^2} \right)$ & 
        $5 \cdot 10^3$ & $5 \cdot 10^{-10}$
        \\

        $ \Theta_1 $ & $(1,4,3/2)_S$ & $y {\Theta_1^{\dagger}} H H^{\dagger} H $ + $ c_{pq} \Theta_1^{\{ijk\}} (L_p^l L_q^m) H^n H^\dagger_l H^o \epsilon_{im} \epsilon_{jn} \epsilon_{ko}$ & $c_{pq} \; \frac{v^3}{\Lambda^3} \left( y \; \frac{v^3}{M^2} \right)$ & 
        $10^2$ & $5 \cdot 10^{-7}$
        \\ \hline \hline

        \end{tabular}
}
\caption{\label{tab:scalar-neutrino-masses-veved}
Same as Tab.~\ref{tab:fermion-neutrino-masses} for the generation of Majorana neutrino masses by the four scalar LSMEs that have a neutral component and may develop a VEV, as explained in Sec.~\ref{sec:scalar-veved}. In the sixth column we quote the upper bound on $\Lambda$ in the limit $\mu \sim M \sim \Lambda$, and in the last column, we display the value of the VEV induced by the SM Higgs doublet.}

\vspace{2.5cm}

\resizebox{\textwidth}{!}{
\begin{tabular}{cccccc}
        \hline \hline
        LSME & $G_{\rm SM} $ & $ \mathcal{L}_{\Delta B = 1}$ & $[L^{S,XY}_{q_1q_2q_3}]_{pqrs}$~\cite{Beneito:2023xbk,Jenkins:2017jig} & Upper Bound (TeV) & $\langle X^0 \rangle$ (TeV) \\
        
        \hline \hline
        
        $ \Xi_1 $ & $(1,3,1)_S$ & $\mu \, \Xi_1^\dagger H H$ + $c_{pq[rs]} \Xi_1 (Q_p L_q) (\bar{d}^\dagger_r \bar{d}^\dagger_s)$ & $[L_{ddu}^{S,RL}]_{pqrs} =c_{rs[pq]}\, \frac{1}{\Lambda^3} \, \left( \mu \; \frac{v^2}{M^2} \right)$ & $5 \cdot 10^5 $ & $6 \cdot 10^{-8}$ \\
        
        $ \Xi$ & $(1,3,0)_S$ & $ \mu \, \Xi H^\dagger H$ + $c_{pq[rs]}  \Xi_{ij}(L^{\dagger}_{pk} \bar{d}^{\dagger}_{q}) (Q_{r}^i Q_{s}^j) H^\dagger_l \epsilon^{kl}$ & $[L_{udd}^{S,LR}]_{pqrs} = c_{rs[pq]} \,\frac{v}{\Lambda^4} \, \left( \mu \; \frac{v^2}{M^2} \right)$ & $ 10^5$ & $3 \cdot 10^{-7}$ \\

        $ \Theta_3 $ & $(1,4,3/2)_S$ & $y \Theta_3 H^\dagger H^\dagger H^\dagger$ + $ c_{[pq]rs}\:\Theta_3^\dagger (Q_p Q_q) (L_r^\dagger \bar u_s^\dagger)$ & $[L_{ddu}^{S,LR}]_{pqrs} =c_{[pq]rs} \, \frac{1}{\Lambda^3} \, \left( y \; \frac{v^3}{M^2} \right)$ & $ 5 \cdot 10^4 $ & $2 \cdot 10^{-8}$  \\

        $ \Theta_1 $ & $(1,4,1/2)_S$ & $y \Theta_1 H^\dagger H H^\dagger$ + $c_{pq[rs]}\:\Theta_1^\dagger (L_p^\dagger \bar d_q^\dagger)(Q_r Q_s)$ & $[L_{udd}^{S,LR}]_{pqrs} =c_{rs[pq]} \, \frac{1}{\Lambda^3} \,\left( y \; \frac{v^3}{M^2} \right)$ &  $ 5 \cdot 10^4$ & $2 \cdot 10^{-8}$ \\
        
        \hline \hline
\end{tabular}
}
\caption{\label{tab:scalar-proton-decay-veved}
Same as Tab.~\ref{tab:fermion-neutrino-masses} for the generation of nucleon decays by the four scalar LSMEs that have a neutral component and may develop a VEV, as explained in Sec.~\ref{sec:scalar-veved}. All the $\Delta B=1$ LEFT WCs lead to the $p\to K^+\nu$ decay channel, from which the limit quoted in the sixth column is obtained.}
}
\end{sidewaystable}


\begin{sidewaystable}[p]
\centering
{
\setlength{\arrayrulewidth}{0.2mm}
\setlength{\tabcolsep}{7pt}
\renewcommand{\arraystretch}{1.9}
\resizebox{\textwidth}{!}{
\begin{tabular}{cccccccc}
         \hline \hline
         LSME & $G_{\rm SM}$ & $\mathcal{L}_{\Delta B = 1}$ & $\Delta d$ &  Op & Matching & Process & Upper Bound (TeV)\\ \hline

        $ Q_7 $ & $ (3,2,7/6)_F$ & $ y_p Q_7^\dagger H \bar u^\dagger_p $ + $ c_{qrs} (\bar Q_7^\dagger \bar e_q^\dagger) (Q_r Q_s) H^\dagger$ & 1 & $\mathcal{O}_{22}^{1111}$ & $C_{qque}^{pqrs} = y_r c_{spq}\left(L + \frac{v^2}{\Lambda^2} \right) \frac{M}{\Lambda} \frac{1}{\Lambda^2} $ & $p \to \pi^{0} e^{+}$ & $ 5 \cdot 10^{11}$ \\

        $ U $ & $ (3,1,2/3)_F$ & $y_p \bar U Q_p H $ + $c_{pqr} (\bar{e}^{\dagger}_{p} \bar{d}^{\dagger}_{q}) (Q_{r} U) H^{\dagger}$ & $1$ & $\mathcal{O}_{22}^{1111}$ & $\mathcal{O}^{qque}_{pqrs} = y_r c_{qps}  \left( L + \frac{v^2}{\Lambda^2} \right) \frac{M}{\Lambda} \frac{1}{\Lambda^2} $ & $p \to \pi^{0} e^{+}$ & $ 5 \cdot 10^{11}$ \\

        $ \Delta_3 $ & $ (1,2,-3/2)_F $ & $ y_p \Delta_3^\dagger H^\dagger \bar e_p^\dagger$ + $ c_{qrs} (\bar \Delta_3^\dagger \bar u_q^\dagger) (Q_r Q_s) H $ & 1 & $\mathcal{O}_{22}^{1111}$ & $C_{qque}^{pqrs} = y_s c_{rpq} \left( L + \frac{v^2}{\Lambda^2} \right) \frac{M}{\Lambda} \frac{1}{\Lambda^2} $  & $p \to \pi^{0} e^{+}$ & $ 5 \cdot 10^{11}$ \\

        $T_2 $ & $ (3,3,2/3)_F$ & $y_p \bar T_2 Q_p H $ + $ c_{qrs} (T_2 Q_q) (\bar u^\dagger_r \bar e^\dagger_s) H^\dagger$ & 1 & $ \mathcal{O}_{22}^{1111}$ &  $C_{qque}^{pqrs} =  y_p c_{qrs} \left( L + \frac{v^2}{\Lambda^2} \right)\frac{M}{\Lambda} \frac{1}{\Lambda^2} $ & $p \to \pi^{0} e^{+}$ & $ 5 \cdot 10^{11}$ \\

         $\Sigma_1 $ & $ (1,3,-1)_F$ & $y_p \bar \Sigma_1 H^\dagger  L_p $ + $ c_{qrs} (\Sigma_1 Q_q) (Q_r Q_s) H $ & 1 & $\mathcal{O}_{18}^{1111}$ & $C_{qqql}^{pqrs} = y_s c_{pqr}\left( L + \frac{v^2}{\Lambda^2} \right) \frac{M}{\Lambda} \frac{1}{\Lambda^2} $ & $p \to \pi^{0} e^{+}$ & $ 5 \cdot 10^{11}$
         \\

         $ N $ & $ (1,1,0)_F $ & $ y_p N^\dagger L_p^\dagger H^\dagger$ + $ c_{qrs}N^\dagger \bar u_q^\dagger \bar d_r^\dagger \bar d_s^\dagger$ & $-1$ & $\mathcal{O}_{10}^{1112}$ & $C_{\bar l dud \tilde H}^{pqrs} = y_p c_{rqs} \frac{1}{M \Lambda^2}$ & $p \to K^{+}\nu$ & $10^8$
         \\

        $T_1 $ & $ (3,3,-1/3)_F$ & $y_p \bar T_1 Q_p H^\dagger $ + $ c_{qrs} (T_1 Q_q) (L^\dagger_r \bar d^\dagger_s)$ & $-1$ & $\mathcal{O}_8^{1112}$ & $C_{\bar l dqq \tilde H}^{pqrs} = y_{r} c_{spq} \frac{1}{M\Lambda^2}$ & $p \to K^{+} \nu $ & $10^8$
        \\

        $ Q_1 $ & $ (3,2,1/6)_F$ &$ y_p Q_1^\dagger H^\dagger \bar u_q^\dagger $ + $ c_{qrs} \bar Q_1^\dagger L_q^\dagger \bar d_r^\dagger \bar d_s^\dagger$ & $-1$ & $\mathcal{O}_{10}^{1112}$ & $C_{\bar l dud \tilde H}^{pqrs} = y_r c_{pqs} \frac{1}{M\Lambda^2}$ & $p \to K^{+} \nu$ & $10^8$
        \\

        $ Q_5 $ & $ (3,2,-5/6)_F$ & $y_p Q_5^\dagger H^\dagger \bar d_p^\dagger$ + $ c_{qrs} (\bar Q_5^\dagger  L_q^\dagger) (Q_r Q_s)$ & $-1$ & $\mathcal{O}_8^{1112}$ & $C_{\bar l dqq \tilde H}^{pqrs} = y_q c_{prs} \frac{1}{M\Lambda^2}$ & $p \to K^{+} \nu$ & $10^8$
        \\

        $ D $ & $ (3,1,-1/3)_F$ &$ y_p \bar D Q_p H^\dagger$ + $ c_{qrs} (D Q_q) (L^\dagger_r \bar d^\dagger_s) $ & $-1$ & $\mathcal{O}_8^{1112}$ & $C_{\bar l d qq \tilde H}^{pqrs} = y_r c_{spq} \frac{1}{M\Lambda^2}$ & $p \to K^{+}\nu$ & $10^8$
        \\

        $\Delta_1 $ & $ (1,2,-1/2)_F$ & $ y_p \Delta_1 H^\dagger \bar e_p$ + $ c_{q[rs]}\: (\bar \Delta_1 Q_q) (\bar d^\dagger_r \bar d^\dagger_s) $ & $-1$ & $\mathcal{O}_9^{1112}$ & $C_{\bar e q dd \tilde H}^{pqrs} = y_p c_{qrs} \frac{1}{M \Lambda^2}$ & $n \to K^{+}e^{-}$ & $10^8$
        \\

        $ \Sigma $ & $ (1,3,0)_F$ & $ y_p \Sigma^\dagger L_p^\dagger H^\dagger$ + $ c_{q[rs]}\:(\Sigma^\dagger \bar d_q^\dagger) (Q_r Q_s) $ & $-1$ & $\mathcal{O}_8^{1121}$ & $C_{\bar l dqq \tilde H}^{pqrs} = y_p c_{q[rs]} \frac{1}{M \Lambda^2}$ & $ n \to K^{+}e^{-}$ & $10^8$ 
        \\
         
        $ E $ & $ (1,1,-1)_F$ & $y_p \bar E^\dagger L_p^\dagger H$ + $ c_{q[rs]}\:(E^\dagger \bar d_q^\dagger) (\bar d_r^\dagger \bar d_s^\dagger)$ & $-1$ & $\mathcal{O}_5^{1112}$ & $C_{\bar l ddd H}^{pqrs} = y_p c_{qrs} \frac{1}{M\Lambda^2}$ & $n \to K^{+}e^{-}$ & $ 5 \cdot 10^7$
        \\

        \hline \hline
\end{tabular}
}
\caption{\label{tab:fermion-proton-decay} Nucleon decay generated by vector-like fermion LSMEs, see also the caption of Tab.~\ref{tab:fermion-neutrino-masses} for details. In this case, colour contractions are generally unique except for the non-renormalisable operators of the colour sextets and octets LSMEs. In such cases, the specific contractions do not affect the quoted limit and are not shown explicitly. SU(2)$_{\rm L}$ indices are not displayed as they do not play a significant role in nucleon decay. Flavour indices $p,q, r, s$ are fixed in the matching onto the SMEFT Wilson coefficients $C_i^{pqrs}$, whereas flavour indices $v, w$ are assumed to be summed over. The fields $N$ and $\Sigma$ are  Majorana fermions.}
}

\end{sidewaystable}

\begin{sidewaystable}[p]
\centering
{
\setlength{\arrayrulewidth}{0.2mm}
\setlength{\tabcolsep}{7pt}
\renewcommand{\arraystretch}{1.9}
\resizebox{\textwidth}{!}{
\begin{tabular}{cccccccc}
         \hline \hline
         LSME & $G_{\rm{SM}} $ & $ \mathcal{L}_{\Delta B = 1}$ & $\Delta d$ &  Op. & Matching & Process & Upper Bound (TeV)\\ \hline

        $\omega_1 $ & $(3,1,-1/3)_S$ & $y_{pq}\omega_1^\dagger (\bar u_p^\dagger \bar e_q^\dagger)$ + $c_{\{rs\}}\: \omega_1 (Q_r Q_s)$ & $-2$ & $\mathcal{O}_2^{1111}$ & $C_{qque}^{pqrs} = y_{rs}c_{pq} \frac{1}{M^2}$ & $p \to \pi^0 e^{+}$ &
        $10^{13}$
        \\

        $\zeta$ & $(3,3,-1/3)_S$ & $y_{pq}\,\zeta^\dagger (Q_p L_q) $ + $c_{[rs]}\:\zeta (Q_r Q_s) $ & $-2$ & $\mathcal{O}_1^{1112}$ & $C_{qqql}^{pqrs} = y_{rs}c_{[pq]} \frac{1}{M^2}$ &  $p \to K^{+}\nu$ &
        $5 \cdot 10^{12}$
        \\

        $\Omega_4$ & $(6,1,4/3)_S$ & $y_{\{pq\}}\: \Omega_4^\dagger (\Bar{u}_p^\dagger \Bar{u}_q^\dagger)$ + $ c_{rstu}\Omega_{4} (\bar{e}^\dagger Q^\dagger) (QQ)  H^\dagger $ & 2 & $\mathcal{O}_{\Omega_{4}}^{111313}$ & $C_{qque}^{pqrs} = y_{r w}c_{s w pq} [y_u]_w\frac{1}{\Lambda^2} \; \epsilon^2$ &  $p \to \pi^{0} e^{+}$ & $5 \cdot 10^{10}$
        \\

        $\varphi$ & $(1,2,1/2)_S$ & $y_{pq} \varphi^\dagger Q_p \bar d_q$ + $c_{rs[tu]} \varphi^i (L^j_r Q^k_s) (\bar d_t^\dagger \bar d_u^\dagger) H^l \epsilon_{il}\epsilon_{jk}$ & 2 & $\mathcal{O}_\varphi^{131131}$ & $C_{qqql}^{pqrs} = y_{w s} c_{sq [vw]} [y_d]_v V^*_{vp}\frac{1}{\Lambda^2}\; \epsilon^2$ & $p \to K^+ \nu$ & $5 \cdot 10^9$
        \\

        $\omega_4$ & $(3,1,-4/3)_S$ & $y_{pq} \omega_4^\dagger (\bar e_p^\dagger \bar d_q^\dagger)$ + $c_{[rs]}\:\omega_4 (\bar u^\dagger_r \bar u^\dagger_s)$ & $-2$ & $\mathcal{O}_3^{1131}$ & $C_{qque}^{pqrs} = y_{s w} c_{[r q]} [y_u]_q [y_d]_w V^*_{w p}\frac{1}{M^2} \; L^\prime$ & $p \to \pi^0 e^{+}$ &
        $5 \cdot 10^8$
        \\

        $\Pi_1$ & $(3,2,1/6)_S$ &  $ y_{pq}\Pi_1^\dagger (L_p^\dagger \Bar{d}_q^\dagger)$ + $c_{rs} \Pi_1 (Q_r Q_s) H^\dagger$ & $-2$ & $\mathcal{O}_8^{1112}$ & $C_{\bar l d qq \tilde H}^{pqrs} = y_{pq} c_{rs} \frac{1}{M^2 \Lambda}$ &  $p \to K^{+}\nu$ & $10^8$
        \\

        $\Pi_7$ & $(3,2,7/6)_S$ & $y_{pq}\Pi_7^\dagger (L_p^\dagger \bar u_q^\dagger)$ + $c_{[rs]}\: \Pi_7 H^\dagger (\Bar{d}^\dagger_r \Bar{d}^\dagger_s) $ & $-2$ & $\mathcal{O}_{10}^{1112}$ & $C_{\bar l d u d \tilde H}^{pqrs} = y_{pr} c_{[qs]} \frac{1}{M^2 \Lambda} $ & $p \to K^{+}\nu$ & $10^8 $
        \\

        $\omega_2$ & $(3,1,2/3)_S$ & $ y_{[pq]}\: \omega_2 (\bar d^\dagger_p \bar d^\dagger_q) $ + $c_{rs}\omega_2^\dagger (L_r^\dagger \bar u_s^\dagger) H^\dagger $ & $-2$ & $\mathcal{O}_{10}^{1112}$ & $ C_{\bar l d u d \tilde H}^{pqrs} = y_{[qs]} c_{pr} \frac{1}{M^2 \Lambda}$ & $p \to K^{+}\nu$ & $10^8 $
        \\

        $\Upsilon$ & $(6,3,1/6)_S$ & $y_{\{pq\}}\: \Upsilon^\dagger
        (Q_p Q_q) $ + $ c_{rstu} \Upsilon (Q_r \bar u_s) (L^\dagger_t \bar d^\dagger_u) $ & 0 & $\mathcal{O}_{40}^{113132}$ & $C_{\bar{l}dqq\tilde{H}}^{pqrs} = y_{\{w s\}}c_{r w p q} [y_u]_w\frac{1}{\Lambda^{3}} \; L$ & $p \to K^{+} \nu$ & $10^7 $
        \\

        $\Phi$ & $(8,2,1/2)_S$ & $y_{pq} \Phi (Q_p \bar u_q) $ + $ c_{rstu} \Phi^\dagger (Q_r Q_s) (L_t^\dagger \bar d_u^\dagger)$ & 0 & $\mathcal{O}_{40}^{111332}$ & $C_{\bar{l}dqq\tilde{H}}^{pqrs} = y_{s w} c_{w r p q} [y_u]_w \frac{1}{\Lambda^3} \; L$ & $p \to K^{+} \nu$ & $10^7$
        \\

        $\Omega_2$ & $(6,1,-2/3)_S$ & $y_{\{pq\}}\: \Omega_2^\dagger (\bar d_p^\dagger \bar d_q^\dagger)$ + $ c_{rstu} \Omega_2 (Q_r \bar u_s)  (L^\dagger_t \bar u^\dagger_u) $ & 0 & $\mathcal{O}_{50}^{131321}$ & $C_{\bar{l}dud\tilde{H}}^{pqrs} = y_{qs} c_{w w p r}[y_u]_w\frac{1}{\Lambda^3} \; L$ & $p \to K^{+} \nu$ & $10^7$
        \\

        $\Xi$ & $(1,3,0)_S$ & $ \mu \, \Xi H^{\dagger}H $ + $c_{pq[rs]} \Xi_{ij}(L^{\dagger}_{pk} \bar{d}^{\dagger}_{q}) (Q_{r}^i Q_{s}^j) H^\dagger_l \epsilon^{kl}$ & 1 & $\mathcal{O}_{45}^{1131}$ & $C_{\bar{l}dqq\tilde{H}}^{pqrs} = c_{p q [r s]} [y_u]_s [y_u]_s \frac{\mu}{\Lambda}\frac{1}{\Lambda^3} \; L^2$ & $n \to K^{+} e^{-}$ & $10^6$
        \\

        $\Xi_1$ & $(1,3,1)_S$ & $ \mu \, \Xi_1^\dagger H H $ + $c_{rs[tu]} \Xi_1 (Q_{r} L_{s}) (\bar{d}^\dagger_{t} \bar{d}^\dagger_{u})$
        & 1 & $\mathcal{O}_{16}^{1132}$ & $C_{qqql}^{pqrs} = c_{q s [vw]} [y_d]_v [y_d]_w V_{vp}^* V_{wr}^* \frac{\mu}{\Lambda} \frac{1}{\Lambda^2} \; L^2$ & $p \to K^{+} \nu$ & $5 \cdot 10^5$
        \\

        $\Theta_3$ & $(1,4,3/2)_S$ & $y \Theta_3 H^{\dagger} H^{\dagger}H^{\dagger}$ + $ c_{[pq]rs}\:\Theta_3^\dagger (Q_p Q_q) (L_r^\dagger \bar u_s^\dagger)$ & 0 & $\mathcal{O}_{37}^{1123}$ & $C_{\bar{l}dqq\tilde{H}}^{pqrs} = y\,c_{[rs]pw} [y_u]_w [y_d]_w V_{w q} \frac{1}{\Lambda^3} \; L^2 $ & $p \to K^{+} \nu$ & $5 \cdot 10^4$
        \\ 

        $\Theta_1$ & $(1,4,1/2)_S$ & $y \Theta_1 H^\dagger HH^\dagger$ + $c_{pq[rs]}\:\Theta_1^\dagger (L_p^\dagger \bar d_q^\dagger)(Q_r Q_s)$ & $0$ & $\mathcal{O}_{45}^{1213}$ & $C_{\bar{l}dqq\tilde{H}}^{pqrs} = y \, c_{pw [v r]} [y_d]_v [y_d]_w V_{w s}^* V_{v q}\frac{1}{\Lambda^3} \; L^2$ & $p \to K^{+} \nu$ & $10^4$
        \\

        $\Omega_1$ & $(6,1,1/3)_S$ & $ y_{[pq]}\: \Omega_1^\dagger (Q_p Q_q) $ + $c_{rstu} \Omega_1 (L^\dagger_r \bar d^\dagger_s)(Q_t \bar u_u) $ & 0 & $\mathcal{O}_{40}^{113132}$ & $C_{\bar{l}dqq\tilde{H}}^{pqrs} = y_{[w r]} c_{pqsw} [y_u]_w\frac{1}{\Lambda^3} \; L$ & $p \to K^{+} \nu$ & $10^7$
        \\

        $\mathcal{S}_1$ & $(1,1,1)_S$ & $y_{[pq]}\:\mathcal{S}_1^\dagger(L_p^\dagger L_q^\dagger)$ + $c_{rstu} \mathcal{S}_1 \bar e^\dagger_p \bar u^\dagger_q \bar d^\dagger_r \bar d^\dagger_s$ & 0 & $\mathcal{O}_{28}^{133121}$  &  $C_{\bar{l}dud\tilde{H}}^{pqrs} =  y_{[pw]}c_{w rqs} [y_e]_w\frac{1}{\Lambda^3} \; L$ & $p \to K^{+} \nu$ & $5 \cdot 10^6 $
        \\

        $\mathcal{S}_2$ & $(1,1,2)_S$ & $ y_{\{pq\}}\:  \mathcal{S}_2^\dagger(\Bar{e}_p \Bar{e}_q) $ + $ c_{rs[tu]}\: \mathcal{S}_2 (\Bar{e}^\dagger_r \Bar{d}^\dagger_s) (\Bar{d}^\dagger_t \Bar{d}^\dagger_u) $ & 0 & $\mathcal{O}_{25}^{133112}$ &  $C_{\bar{e}dddD}^{pqrs} = y_{\{p w\}}c_{w q r s} \frac{1}{\Lambda^3} \; L$  & $n \to K^{+} e^{-}$ &
        $10^6$
        \\

        \hline \hline
\end{tabular}
}
\caption{\label{tab:scalar-proton-decay}
Same as Tab.~\ref{tab:fermion-proton-decay} for nucleon decays induced by scalar LSMEs. 
}
}
\end{sidewaystable}

\begin{sidewaystable}[p]
\centering
{
\setlength{\arrayrulewidth}{0.2mm}
\setlength{\tabcolsep}{7pt}
\renewcommand{\arraystretch}{1.9}
\resizebox{\textwidth}{!}{
\begin{tabular}{cccccccc}
         \hline \hline

         LSME & $G_{\rm SM}$ & $ \mathcal{L}_{\Delta B = 1}$ & $\Delta d$  & Op. & Matching & Process & Upper Bound (TeV)\\ \hline

        $ \mathcal{Q}_5 $ & $ (3,2,-5/6)_V$ & $y_{pq} \mathcal{Q}_5^\dagger \bar e_p^\dagger Q_q$ + $c_{rs} \mathcal{Q}_5 Q_r \bar u^\dagger_s$ & $-2$ & $\mathcal{O}_2^{1111}$ & $C_{qque}^{prst} = y_{tp} c_{rs} \frac{1}{M^2}$ & $p \to \pi^0e^{+}$ & 
        $10^{13}$
        \\

        $\mathcal{Q}_1 $ & $ (3,2,1/6)_V$ & $y_{pq} \mathcal{Q}_1^\dagger \bar u_p^\dagger L_q$ + $ c_{rs} \mathcal{Q}_1 Q_r \bar d^\dagger_s $ & $-2$ & $\mathcal{O}_4^{1111}$ & $C_{duql}^{pqrs} = y_{qs} c_{rp} \frac{1}{M^2}$ & $p \to \pi^0 e^{+}$ &
        $5 \cdot 10^{12}$
        \\

        $\mathcal{L}_1 $ & $ (1,2,1/2)_V$ & $\mu\: \mathcal{L}_1^\dagger DH$ + $c_{pqrs} \mathcal{L}_1 L_p \bar d^\dagger_q \bar d^\dagger_r \bar u^\dagger_s$ & 1 & $\mathcal{O}_{12}^{1131} $ & $C_{duql}^{pqrs} = c_{s w p q}[y_d]_r V_{r w}^* \frac{\mu}{\Lambda} \frac{1}{\Lambda^2} \; L$  & $p \to K^{+} \nu$ & $5 \cdot 10^9$
        \\

        $\mathcal{U}_2 $ & $ (3,1,2/3)_V$ & $y_{pq}\: \mathcal{U}_2^\dagger Q_p  L_q^\dagger $ + $ c_{rs} \: \mathcal{U}_2 \Bar{d}^\dagger_r Q_s H^\dagger $ & $-2$ & $\mathcal{O}_8^{1112}$ & $C_{\bar l d qq \tilde H}^{pqrs} = y_{rp} c_{qs} \frac{1}{M^2 \Lambda}$ & $p \to K^{+}\nu$ & $10^8$
        \\

        $\mathcal{X} $ & $ (3,3,2/3)_V$ & $ y_{pq}\mathcal{X}^\dagger Q_p L_q^\dagger$ + $ c_{rs} \mathcal{X} \Bar{d}^\dagger_r Q_s H^\dagger $ & $-2$ & $\mathcal{O}_8^{1112}$ & $C_{\bar l d qq \tilde H}^{pqrs} = y_{rp} c_{qs} \frac{1}{M^2 \Lambda}$ & $p \to K^{+}\nu$ & $10^8$
        \\

        $ \mathcal{Y}_1 $ & $ (\bar 6,1,1/6)_V$ &$ y_{pq} \mathcal{Y}_1^\dagger Q_p^\dagger \bar d_q $ + $ c_{rstu} \mathcal{Y}_1 Q^\dagger_r Q^\dagger_s \bar u^\dagger_t L_u $ & 0 & $\mathcal{O}_{40}^{113132}$ & $C_{\bar{l}dqq\tilde{H}}^{pqrs} = y_{w q} c_{s r w p} [y_u]_w \frac{1}{\Lambda^3} \; L$ & $p \to K^{+} \nu$ & $10^7$
        \\

        $ \mathcal{Y}_5 $ & $ (\bar 6 ,1,-5/6)_V$ &$ y_{pq} \mathcal{Y}_5  Q_p \bar u^\dagger_q  $ + $ c_{rstu} \mathcal{Y}_5^\dagger \bar u_r \bar d_s^\dagger \bar d_t^\dagger L_u^\dagger $ & 0 & $\mathcal{O}_{50}^{131321}$ & $C_{\bar{l}dud\tilde{H}}^{pqrs} = y_{w r} c_{w q sp} [y_u]_w \frac{1}{\Lambda^3} \; L $ & $p \to K^{+} \nu$ & $10^7$
        \\

        $\mathcal{G}_1 $ & $ (8,1,1)_V$ & $y_{pq}\mathcal{G}_1 \Bar{u}_p\bar d^\dagger_q $ + $ c_{rstu} \mathcal{G}_1^\dagger Q_r (Q_s Q_t) L_u^\dagger$ & 0 & $\mathcal{O}_{40}^{111332}$ & $C_{\bar{l}dqq\tilde{H}}^{pqrs} = y_{w q} c_{rs w p} [y_u]_w \frac{1}{\Lambda^3} \; L$ & $p \to K^{+} \nu$ & $10^7$
        \\

        $\mathcal{G} $ & $ (8,1,0)_V$ & $ y_{pq}\mathcal{G}\bar u^\dagger_p \bar u_q $ + $c_{rstu} \mathcal{G} Q_r L^\dagger_s \bar d^\dagger_t \bar d^\dagger_u $ & 0 & $\mathcal{O}_{50}^{131321}$ & $C_{\bar{l}dud\tilde{H}}^{pqrs} = y_{r w} c_{w pqs}[y_u]_w \frac{1}{\Lambda^3} \; L$ & $p \to K^{+} \nu$ & $10^7$
        \\

        $\mathcal{B} $ & $ (1,1,0)_V$ & $y_{pq}\mathcal{B}\bar u^\dagger_p \bar u_q$ + $c_{rstu} \mathcal{B}Q_r L^\dagger_s \bar d^\dagger_t \bar u^\dagger_u$ & 0 & $\mathcal{O}_{50}^{131321}$ & $C_{\bar{l}dud\tilde{H}}^{pqrs} = y_{r w} c_{w pqs} [y_u]_w \frac{1}{\Lambda^3}\; L$ & $p \to K^{+} \nu$ & $10^7$
        \\

        $\mathcal{B}_1 $ & $ (1,1,1)_V$ & $ y_{pq}\mathcal{B}_1 \bar u_p \bar d^\dagger_q $ + $ c_{rstu}\mathcal{B}_1^\dagger Q_r (Q_s Q_t) L_u^\dagger$ & 0 & $\mathcal{O}_{40}^{131132}$  & $C_{\bar{l}dqq\tilde{H}}^{pqrs} = y_{w q} c_{w rsp} [y_u]_w \frac{1}{\Lambda^3} \; L $ & $p \to K^{+} \nu$ & $10^7$
        \\

        $\mathcal{H} $ & $ (8,3,0)_V$ & $y_{pq}\mathcal{H} Q^\dagger_p Q_q$ + $ c_{rstu} \mathcal{H} Q_r \bar d^\dagger_s \bar d^\dagger_t L^\dagger_u$ & 0 & $\mathcal{O}_{42}^{111332}$ & $C_{\bar{l}dqq\tilde{H}}^{pqrs} = y_{w r} c_{s v q p}[y_d]_w V_{w v} \frac{1}{\Lambda^3}\; L$  & $p \to K^{+} \nu$ & $5 \cdot 10^6$
        \\

        $\mathcal{W} $ & $ (1,3,0)_V$ & $ y_{pq}\mathcal{W}Q^\dagger_p Q_q$ + $c_{rstu} \mathcal{W} Q_r L^\dagger_s \bar d^\dagger_t \bar d^\dagger_u$ & 0 & $\mathcal{O}_{42}^{111323}$ & $C_{\bar{l}dqq\tilde{H}}^{pqrs} = y_{v r} c_{spq w} [y_d]_v V_{v w} \frac{1}{\Lambda^3} \; L $ & $p \to K^{+} \nu$ & $5 \cdot 10^6$
        \\

        $\mathcal{W}_1 $ & $ (1,3,1)_V$ & $y\mathcal{W}_1 H^\dagger DH^\dagger$ +  $c_{pqrs} \mathcal{W}_1^\dagger Q_p L_q^\dagger \bar d_r^\dagger \bar u_s^\dagger$ & 0 & $\mathcal{O}_{44}^{1131}$ & $C_{\bar{l}dqq\tilde{H}}^{pqrs} = y \, c_{rpqs} [y_u]_{s} \frac{1}{\Lambda^3} \; L$  & $p \to K^{+} \nu$ & $ 5 \cdot 10^6$
        \\

        $\mathcal{L}_3 $ & $ (1,2,-3/2)_V$& $y_{pq}\mathcal{L}_3 \bar e_p L^\dagger_q$ + $c_{rstu} \mathcal{L}_3^\dagger \bar e_r^\dagger Q_s \bar d_t^\dagger \bar d_u^\dagger$ & 0 & $\mathcal{O}_{49}^{331121}$ & $C_{\bar{e}qdd\tilde{H}}^{pqrs} =  y_{p w} c_{w qrs} [y_e]_w \frac{1}{\Lambda^3} \; L $ & $n \to K^{+} e^{-}$ & $ 5 \cdot 10^6$
        \\

        $\mathcal{U}_5 $ & $ (3,1,5/3)_V$ & $ y_{pq} \mathcal{U}_5^\dagger \bar e_p  \bar u_q^\dagger$ + $ c_{rstu} \mathcal{U}_5 \bar d^\dagger_r (\bar d^\dagger_s \bar d^\dagger_t) \bar u_u$ & 0 & $\mathcal{O}_{33}^{122121}$ & $C_{\bar{e}dddD}^{pqrs} = y_{p w} c_{qrs w} \frac{1}{\Lambda^3} \; L $ & $n \to K^{+} e^{-}$ & $10^6$
        \\
        
        \hline \hline
\end{tabular}
}
\caption{\label{tab:vector-proton-decay}
Same as Tab.~\ref{tab:fermion-proton-decay} for nucleon decays induced by vector LSMEs. Explicit Lorentz contractions involving sigma matrices $\sigma$ or $\bar \sigma$ are not shown, as they are uniquely determined for renormalisable operators but remain ambiguous for non-renormalisable ones. However, up to $\mathcal{O}(1)$ factors this ambiguity does not affect the limits quoted in the last column.}
}
\end{sidewaystable}


\bibliography{main}

\end{document}